\documentclass[11pt]{article}

\usepackage{putex}

\RequirePackage[a4paper,top=30.6mm,bottom=38.6mm,left=34.6mm,right=34.6mm,footskip=1.3cm]{geometry}
\usepackage{setspace}
\usepackage{booktabs}
\usepackage{cancel}
\usepackage{multirow}
\usepackage{comment}
\usepackage{bbold}
\usepackage{caption}
\usepackage{amsmath}
\usepackage{enumerate}
\usepackage{cite}
\usepackage{tensor}
\usepackage{slashed}
\usepackage[utf8]{inputenc}
\usepackage{rotating}
\usepackage{bigfoot}
\usepackage[
colorlinks=true,
linkcolor=black,
urlcolor=blue,
filecolor=black,
citecolor=red,
]{hyperref}

\usepackage[textsize=footnotesize]{todonotes}

\numberwithin{equation}{section}

\def \< {\left<}
\def \> {\right>}

\newcommand{\be}{\begin{equation}} \newcommand{\ee}{\end{equation}}
\newcommand{\bea}{\begin{eqnarray}}  \newcommand{\eea}{\end{eqnarray}}
\newcommand{\nn}{\nonumber}

\usepackage{amsfonts, amsthm}
\usepackage[english]{babel}
\usepackage{slashed}
\usepackage{mathrsfs}
\usepackage{amssymb}
\usepackage{color}

\newcommand{\mrm}[1]{\mbox{$\mathrm{#1}$}}

\begin{document}

\begin{flushright}
\hfill{MPP-2023-113}
\end{flushright}
\vspace{0.5cm}

	\begin{center}        
		\huge Towards AdS Distances in String Theory
	\end{center}
	
	\vspace{0.7cm}
	\begin{center}        
		{\large  Yixuan Li$^1$,\;\; Eran Palti$^2$,\;\; Nicol\`o Petri$^2$, }
	\end{center}
	
	\vspace{0.15cm}
	\begin{center}  
	\emph{${}^1$ Max-Planck-Institut f\"ur Physik (Werner-Heisenberg-Institut),
             F\"ohringer Ring 6,
             80805, M\"unchen, Germany}\\[.2cm]
		\emph{${}^2$ Department of Physics, Ben-Gurion University of the Negev, Be'er-Sheva 84105, Israel}\\[.3cm]
		\emph{}\\[.2cm]
		e-mails:  \tt yixuan@mpp.mpg.de \; palti@bgu.ac.il, \;petri@post.bgu.ac.il
	\end{center}
	
	\vspace{1cm}
	
	
	\begin{abstract}
	\noindent  
	The AdS Distance Conjecture proposes to assign a notion of distance between AdS vacua in quantum gravity. We perform some initial developments of this idea. We first propose more sharply how to define a metric on conformal variations of AdS through the action. This metric is negative, making the distance ill-defined, a property relating to the famous conformal factor problem of quantum gravity. However, in string theory, variations of the AdS conformal factor are accompanied by variations of the internal dimensions and of the background flux. We propose an {\it action metric}, which accounts for all of these variations simultaneously. Accounting for the variations of the overall volume of the internal dimensions can flip the sign of the action metric making it positive. This positivity is related to the absence of scale separation between the internal and external dimensions: the negative external conformal contribution must be sub-dominant to the positive internal contribution. We then focus specifically on the families of solutions of eleven-dimensional supergravity on AdS$_4 \times S^7$ and AdS$_7 \times S^4$. For these, there is only a single further additional contribution to the action metric coming from variations of the Freund-Rubin flux. This contribution is subtle to implement, and the unique prescription we find requires singling out the radial direction of AdS as special. Adding the flux contribution yields an overall total action metric which becomes positive for both the AdS$_4$ and AdS$_7$ families of solutions. The final result is therefore a procedure which yields a well-defined distance for these families of solutions. 
	\end{abstract}
	
	\thispagestyle{empty}
	\clearpage
	
	\tableofcontents
	
	\setcounter{page}{1}

\section{Introduction}
\label{sec:intro}

Moduli spaces in compactifications of string theory parameterise deformations of the metric of the internal manifold, as well as deformations of other fields, such as the Ramond-Ramond forms. The moduli space has a metric on it, and therefore a notion of distance. It has been conjectured that infinite distance variations in moduli spaces correspond to a certain breakdown of the effective theory due to a tower of states becoming light \cite{Ooguri:2006in}. There is significant evidence for this proposal by now, see \cite{Palti:2019pca,vanBeest:2021lhn} for reviews, and \cite{Baume:2016psm,Klaewer:2016kiy,Blumenhagen:2018nts,Grimm:2018ohb,Grimm:2018cpv,Corvilain:2018lgw,Lee:2018spm,Lee:2019wij} for some of the initial work. 

The metric on the moduli space can be derived by varying the higher-dimensional Einstein-Hilbert action. There are two types of important variations of the metric of the compact internal manifold: infinitesimal transverse-traceless deformations of the metric, and the overall conformal factor (volume). We can focus on the simpler conformal rescaling of the internal manifold, which we take to be Calabi-Yau for simplicity. The metric on the space of conformal variations of the Calabi-Yau can be extracted by allowing the Calabi-Yau volume to depend on the external manifold coordinates, and extracting the term in the action which is quadratic in external derivatives acting on the Calabi-Yau volume. If we denote the volume as $\sigma$, the ten-dimensional metric as $g$ and Ricci scalar as $R$, and the four-dimensional (Einstein frame) metric as $g_4$ and Ricci scalar as $R_4$, then we can write
\be
\frac{1}{2 \kappa^2}\int_{{\cal M}_4\times CY} \sqrt{g}\; R \supset \frac12 \int_{{\cal M}_4} \sqrt{g_4} \Big[ R_4 - K_{\sigma\sigma}\left(\partial \sigma\right)^2\Big] \;. 
\label{volcy}
\ee 
The metric on the space of volumes is then $K_{\sigma\sigma}$. Importantly, this metric is positive definite $K_{\sigma\sigma}>0$. We therefore have a good notion of the proper distance $\Delta$ as measured along a geodesic path $S$, with line elements $d\tau$:
\be
\Delta = \int_S \left( K_{\sigma\sigma} \frac{\partial \sigma}{\partial \tau}\frac{\partial \sigma}{\partial \tau}\right)^{\frac12} d\tau \;.
\label{disvolcy}
\ee

It is important to note that in order to extract the metric on the space of volumes through (\ref{volcy}), we had to give the volume some arbitrary dependence on the external spacetime coordinates, so as to extract the coefficient in front of its derivatives. The ten-dimensional solution, of Minkowski space times a Calabi-Yau, is not a solution if the Calabi-Yau volume actually depends on the four-dimensional coordinates. This would not be a product metric. Once we extracted the metric on volumes from the kinetic term, we can now measure the distance along variations of the volumes in which the volume is constant in spacetime. Those constant variations are solutions of the equations of motions. So we extract a notion of distance between different solutions of string theory. 

In \cite{Lust:2019zwm}, it was suggested that one could define a similar notion of distance between solutions of string theory also for solutions which are not of the type Minkowski times Calabi-Yau. In particular, one could consider distances between Anti-de Sitter (AdS) solutions in string theory. The simplest notion of this is distances between AdS vacua that are part of infinite families of solutions in which the AdS radius varies. 

It was noted that the distance on the volume variations of the internal manifold (\ref{disvolcy}) (as well as on the other deformations) can be derived by using a de-Witt type formula on variations in the space of metrics \cite{DeWitt,CANDELAS1991455}
\be
\Delta = \int_S \left(\frac{1}{\sigma} \int_{CY} \sqrt{h}\; h^{mn} h^{op} \frac{\partial h_{mo}}{\partial \tau}\frac{\partial h_{np}}{\partial \tau} \right)^{\frac12} d\tau \;.
\label{cydimredfor}
\ee 
Here $h_{mn}$ stands for the metric on the internal manifold, say the Calabi-Yau, and $\sigma$ its volume. One could then consider using a similar formula to extract the distance on the AdS conformal factors, by replacing the internal metric $h$ with the external (AdS) metric $g$. This yields a divergence approaching the flat space limit of AdS, which was proposed to be identified with an infinite distance \cite{Lust:2019zwm}.

The formula (\ref{cydimredfor}) arises simply from dimensional reduction of the internal part of the higher-dimensional Einstein-Hilbert action.\footnote{More precisely, it comes from a mixed component since we have external derivatives acting on the internal metric.} And so it was proposed in \cite{Lust:2019zwm} that similarly one could extract the distance on external variations from the external part of the higher-dimensional Einstein-Hilbert action. But it was not clear how to do this precisely, specifically relating to what is the role of the Einstein frame condition. One thing which was clear is that the metric extracted this way would be negative. A negative metric does not lead to a well-defined notion of proper distance, and so this is problematic.

We note that this negativity of the metric on conformal variations of the spacetime metric is related to a famous and old problem in Euclidean quantum gravity called the conformal factor problem (see for instance \cite{Gibbons:1976ue,Marolf:2022ybi}). The negative kinetic terms of the conformal factor mean that a path-integral interpretation of quantum gravity suffers from runaway solutions, and is ill-defined. This problematic negative behaviour is therefore very deep and old.

It was suggested in \cite{Lust:2019zwm} that it may be that the negativity of the metric can be cured by noting that in string theory families of AdS solutions vary not only the conformal factor of AdS, but simultaneously also other quantities, such as the internal dimensions and various fluxes. The notion of distance between solutions must account for these, leading to a type of Generalized Distance Conjecture.  It was proposed that the metric on all the variations, something which we denote here as the {\it action metric}, will be overall positive. 

The proposed prescription for calculating the action metric was: take a family of solutions parametrised by some parameter $\sigma$, give this parameter an infinitesimal spacetime dependence, extract the coefficient in front of the two-derivative terms in the action acting on $\sigma$, and use this as the metric on the space of solutions parameterised by $\sigma$. 

In this paper we initiate an exploration of this concept of action metrics. 
Our aim is to try and follow the procedure suggested in \cite{Lust:2019zwm}, and described above, to see what the result is. We encounter various issues and subtleties in trying to implement this, and propose solutions to them. At the end, we do find a well-defined notion of an action metric, that can be calculated through our procedures explicitly, both for the specific solutions we study and more general solutions. 

However, this is really an exploratory investigation, we do not know if the procedures we propose are correct, and more generally, if the whole notion of the distance between solutions makes sense. We therefore view our results as first steps towards understanding this notion of distance.

The simplest sets of solutions of string theory in which to try and calculate the action metric, and so distance, are those which depend only on one parameter. So the distance is measured along the real line, and the metric is a scalar which, by appropriate coordinate choices, can be made a constant. Explicitly, we consider the simplest one-parameter set of solutions, those of Freund-Rubin type. The original Freund-Rubin solutions are the one-parameter infinite family of solutions AdS$_4 \times S^7$ and AdS$_7 \times S^4$ of eleven-dimensional supergravity. 

The results we find, using our proposed prescription for the action metric, is as follows: The variation of only the conformal factor of AdS leads to a negative metric on the family of solutions. Adding to this the variation of the internal space can make the metric positive. In general, this happens if there is no scale separation, so the AdS radius scales in the same way as the internal volume with the parameter of the solutions, and if the dimension of the AdS factor is less than six. Since the Freund-Rubin solutions do not have scale separation, this yields a positive metric on the AdS$_4$ family of solutions, and a negative metric on the AdS$_7$ family. The final contribution to the action metric should come from the fluxes. Calculating this contribution is very subtle. We identify a procedure which seems to lead to a well-defined contribution, but relies on picking out the radial direction of AdS specifically. With our procedure, we find that the contribution to the action metric from the fluxes is positive, and yields a final total action metric which is positive for both the AdS$_4$ and AdS$_7$ families of solutions. The positive metric then implies a well-defined distance. 

Let us note that there has been work on trying to understand the distance notions of \cite{Lust:2019zwm}, in various settings. See \cite{Rudelius:2021oaz,Basile:2021mkd,Angius:2022aeq,Montero:2022ghl,Farakos:2023nms,Buratti:2020kda,Luben:2020wix,Li:2021utg,Collins:2022nux,Cribiori:2022trc,Shiu:2022oti,Cribiori:2023swd} for an incomplete list. Our approach is significantly different to these works, but it may be that there are some relations.

The paper is set out as follows. In section \ref{sec:genactiondistance} we study the metric on variations of the (spacetime) metric, including both internal and external variations. In section \ref{sec:metflux} we study the contribution to the metric coming from the flux variations, focusing on the case of Freund-Rubin solutions. We present a summary and discussion in section \ref{sec:dis}. 

A note on conventions: we work in Planck units $M_p=1$, where the Planck scale is the lower ($d$-dimensional one). The higher ($D$-dimensional) gravitational constant is denoted as $\kappa^2_D$.
	
\section{The metric on metric variations}
\label{sec:genactiondistance}
	
In this section we introduce a notion of metric over the space of metric variations starting from the study of the Einstein-Hilbert action. We consider first metric variations which are given by Weyl rescalings of the AdS factor. We calculate the metric over the space of such variations and we show that it is negative definite. This follows the ideas of \cite{Lust:2019zwm}, but we sharpen and clarify the procedure for obtaining the metric. Next we consider the contribution to the metric coming from variations of the internal volume. We then combine the two results obtaining a general formula including both external and internal metric variations. We finally apply our general result to the specific case of Freund-Rubin AdS vacua in string theory.

\subsection{External metric variations}
\label{externalvariations}

Perhaps the simplest examples of metric variations are Weyl rescalings. Given a homogeneous Einstein space $M_d$, one can consider the following deformation of the metric
\begin{equation}\label{externalvariations}
 ds_d^2=e^{2\sigma}\,d\hat s^2_d\,,
\end{equation}
where $d\hat s^2_d$ is the metric defined on $M_d$ and the conformal mode $\sigma$ is, in general, a function defined over $M_d$. 

In the case where $\sigma$ is constant, this deformation relates two AdS spaces of different cosmological constants:
\begin{equation}
\begin{split}
 \frac{\Lambda}{\hat{\Lambda}}=e^{-2\sigma}\,.
 \end{split}
\end{equation}
This type of constant rescaling relates different members of families of solutions in string theory (or M-theory), such as AdS$_4 \times S^7$. 

A way to define a notion of distance in such a family of solutions is to promote $\sigma$ to have spacetime dependence and extract the resulting kinetic terms. We call the procedure of giving $\sigma$ spacetime dependence `` {\it gauging $\sigma$} ", in the sense that it promotes a global parameter to a local one. Performing such a gauging implies that the equations of motion are no longer solved, since the space is no longer AdS. We will refer to this as going `` {\it off-shell} ". This is true also of the better understood moduli space metrics. At the end, once the metric is extracted, we can consider constant variations of $\sigma$, like we do with moduli expectation values, which are back `` {\it on-shell} ".

Let us derive such an off-shell metric for Weyl rescalings of the type \eqref{externalvariations}. The kinetic terms for $\sigma$ arise from the Einstein-Hilbert term $\sqrt{-g_d}\,R_d$, and can be easily determined through the usual transformation of the Ricci scalar $R_d$ under Weyl rescalings,
\begin{equation}\label{WeylTransform}
 R_d=e^{-2\,\sigma}\,\left(\hat R_d-(d-1)(d-2)\hat g_d^{mn} \partial_m \,\sigma \partial_n \,\sigma-2(d-1)\,\hat \nabla_d^2\,\sigma   \right)\,,
\end{equation}
where the hatted quantities are defined by contractions with the undeformed metric $d\hat s^2_d$ on $M_d$. 
The laplacian $\hat \nabla_d^2$ can be written in the usual Laplace-Beltrami form
\begin{equation}\label{laplacian}
 \hat \nabla_d^2\,\sigma=\frac{1}{\sqrt{-\hat g_d}}\,\partial_m\,\left(\sqrt{-\hat g_d}\,\hat g_d^{mn}\,\partial_n\,\sigma    \right)\,.
\end{equation}
We can thus derive the action metric starting from \eqref{WeylTransform} and performing a partial integration of the $\hat \nabla_d^2$ term,
\bea
\label{EHtermExternalVolumeIntByPart}
 S_d &=&\frac12\int d^dx\sqrt{-g_d}\,R_d=\frac12\int d^dx\sqrt{-\hat g_d}\,e^{(d-2)\,\sigma}\,\left(\hat R_d-(d-1)(d-2)\hat g_d^{mn} \partial_m \,\sigma \partial_n \,\sigma \right) \nn \\
 & &+(d-1)\,\int d^dx\sqrt{-\hat g_d}\,\partial_m \left(e^{(d-2)\,\sigma}\right)\,\hat g_d^{mn}\,\partial_n\sigma \nn \\
 &=&\frac12\int d^dx\sqrt{-\hat g_d}\,e^{(d-2)\,\sigma}\,\biggl(\hat R_d -(d-1)(d-2)\hat g_d^{mn} \partial_m \,\sigma \partial_n \,\sigma \nn  \\
 &+&2(d-1)(d-2)\hat g_d^{mn} \partial_m \,\sigma \partial_n \,\sigma \biggr)\,,
\eea
where we crucially included in the partial integration the volume factor $e^{d\sigma}$ and we omitted the total derivative.

At this point there was an ambiguity in \cite{Lust:2019zwm} regarding how to interpret this kinetic term. We propose that the natural interpretation, raised as a possibility already in \cite{Lust:2019zwm}, is that one absorbs the Weyl factor back into the metric (and Ricci scalar). 
The Einstein-Hilbert term can be thus written as,
\begin{equation}
\begin{split}\label{EHtermExternalVolume}
 S_d=&\,\frac12\int d^dx \sqrt{-g_d}\,\left(\tilde{R}_d+(d-1)(d-2)(\partial \,\sigma)^2 \right)\\
  &=\frac12\int d^dx\sqrt{-g_d}\,\left(\tilde{R}_d-K_{\sigma\sigma}(\partial \,\sigma)^2 \right)\,,\\
  \end{split}
\end{equation}
where we stress that we absorbed the factor $e^{d\sigma}$ within the volume factor $ \sqrt{-g_d}$ and used that $\hat g_d^{mn} e^{-2\sigma}=g_d^{mn}$ in the kinetic term for $\sigma$.\footnote{In \eqref{EHtermExternalVolume} we use the notatiom $(\partial \,\sigma)^2 $ to indicate that the contraction is performed through the deformed metric $ds^2_d=e^{2\sigma}\,\,d\hat s^2_d$.}  We also defined $\tilde{R}_d$ as $R_d$ for a constant $\sigma$, so
\be
\tilde{R}_d = e^{-2\,\sigma}\hat R_d \;.
\ee
The Einstein frame is then such that the Planck mass is defined as the prefactor of $\tilde{R}_d$.
From the above expression we can extract the metric over the space of $\sigma$-variations as the coefficient $K_{\sigma\sigma}$ of the kinetic term:
\begin{equation}\label{externaldistance}
 K_{\sigma\sigma}=-(d-1)(d-2)\,.
\end{equation}
Note that this is negative. 

\subsection{Internal metric variations}
\label{internalvariations}

We have discussed the contribution to the metric from the external AdS conformal factor variation. In this section we consider the metric on variations of the volume of the internal dimensions. We keep the external AdS constant in this part, and subsequently will combine the two types of variations.

We consider a product geometry $M_d\times Y_k$ where $Y_k$ is a compact manifold and $M_d$ being the external spacetime. Internal volume variations are described by a modulus $\tau$ defined over $M_d$ as 
\begin{equation}\label{internalvariationmetric}
  ds_D^2=d s^2_d+e^{2\tau}\,d\hat s^2_k\,,
\end{equation}
where $d\hat s^2_k$ is the metric $Y_k$.
The computation of the metric on volume variations of the internal manifold is completely standard. We reproduce it here for completeness. We compactify the $D$-dimensional action to $d$ dimensions and secondly cast the action into the Einstein frame. 

We start by expanding the $D$-dimensional Ricci scalar with the following prescription,
\begin{equation}\label{Rdimreduction}
 R= R_d+e^{-2\tau}\, R_k-k(k+1)\,  g_d^{mn} \partial_m \,\tau \partial_n\tau-2k\, \nabla_d^2\,\tau\,,
\end{equation}
where $ \nabla_d^2$ was defined in \eqref{laplacian}. We derive the above reduction formula in appendix \ref{app:reduction}. Since the modulus $\tau$ depends only on the external coordinates, we express the contractions with $\tau$-derivatives directly in terms of $d$-dimensional indices $(m,n, \dots)$.
As in the previous section, we proceed by integrating by parts the Laplacian $\nabla_d^2\tau$ including the volume factor in the derivative,
\bea
 \label{internaleh}
  S&=&\frac{1}{2\kappa_D^2}\int d^Dx\sqrt{-g}\,R=\frac{1}{2\kappa_D^2}\int d^{D}x\sqrt{\hat g_k}\,\sqrt{- g_d}\,e^{k\,\tau}\,\bigl( R_d+e^{-2\tau}\,\hat R_k\\
  &-&k(k+1)\,g_d^{mn} \partial_m \,\tau \partial_n\tau \nn 
  +2k\,e^{-k\tau}\,\partial_m\bigl(e^{k\,\tau}\bigr)\, g_d^{mn}\partial_n\tau\bigr) \nn\\
  &=&\frac{1}{2\kappa_D^2}\int d^Dx\sqrt{\hat g_k}\,\sqrt{- g_d}\,e^{k\,\tau}\,\left( R_d+e^{-2\tau}\,\hat R_k+k(k-1)\, g_d^{mn} \partial_m \,\tau \partial_n\tau \right).
\eea
We need now to pass to the $d$-dimensional Einstein frame. To do so we first perform a Weyl rescaling within $M_d$ introducing the metric $d s_{E}^2$ as 
\begin{equation}
 \begin{split}\label{EinteinFrameTransform}
  &d s_{d}^2= e^{2\omega}\,ds_{E}^2\,\\
  & R_d= e^{-2\omega}\,\left(R_{E}-(d-1)(d-2)g_{E}^{mn} \partial_m \omega \,\partial_n\omega-2(d-1)\,\nabla_E^2\,\omega   \right)\,,
 \end{split}
\end{equation}
where $\nabla_E^2\,\omega$ is expressed in terms of the metric $d s^2_{E}$. After this rescaling the Einstein-Hilbert action takes the form
\begin{equation}
 \begin{split}\label{ehterm}
  S=\frac{1}{2\kappa_D^2}\int d^Dx\sqrt{\hat g_k}\sqrt{- g_{E}}\,e^{(d-2)\omega}&\,e^{k\,\tau}\,\bigl( R_{E}+e^{-2\tau+2\omega}\,\hat R_k+k(k-1)\, g_{E}^{mn} \,\partial_m \tau \partial_n\tau \,\\
  &-(d-1)(d-2) g_{E}^{mn} \,\partial_m \omega \partial_n\omega-2(d-1)\,\nabla_E^2\,\omega \bigr)\,.
 \end{split}
\end{equation}
The $d$-dimensional Einstein frame is defined by the following choice
\begin{equation}\label{EinsteinFrameCondition}
 \omega=-\frac{k}{d-2}\,\tau\,.
\end{equation}
Imposing the Einstein frame condition \eqref{EinsteinFrameCondition} into the action \eqref{ehterm}, the laplacian $\nabla_E^2\,\omega $ becomes a total derivative. This leads to the following $d$-dimensional action\footnote{In the reduction we fixed $\frac{\text{Vol}(Y_k)}{\kappa_D^2}=1$ and we used the notation $(\partial \,\tau)^2= g_{E}^{mn}\partial_m \tau \partial_n\tau$.},
\begin{equation}\label{EinsteinFrameActioninternal}
 \begin{split}
  S_d=&\,\frac12\int d^dx\,\sqrt{-g_{E}}\,\bigl( R_{E}+e^{-2\tau+2\omega}\,\hat  R_k-k^2\,\left(\frac{d-1}{d-2}-\frac{k-1}{k}\right)\,( \partial \,\tau)^2 \bigr)\\
  &=\frac{1}{2}\int d^dx\,\,\sqrt{- g_{E}}\,\left(R_{E}+e^{-2\tau+2\omega}\,\hat R_k-K_{\tau\tau}\,( \partial \,\tau)^2 \right)\,.
 \end{split}
\end{equation}
In analogy to what we did in previous section we can extract the contribution to the action metric from internal volume variations looking at the kinetic term of $\tau$,
\begin{equation}\label{internaldistance}
K_{\tau\tau}=k^2\,\left(\frac{d-1}{d-2}\,-\frac{k-1}{k}\right)\,.
\end{equation}

\subsection{Combined metric variations}
\label{genvariationsmetric}

In this section we consider combining both external conformal and internal volume variations. For constant variations, it is clear that the two variations are independent and so can be added straightforwardly. When they are `gauged', so given some spatial dependence, it is less obvious but nonetheless true that they remain commuting. We show this explicitly in appendix \ref{app:extandint}. The final result for the total metric variations is then:
 \begin{equation}
 \label{genvariations2mt}
 \begin{split}
  S_d=\frac12\int d^dx\,\,\sqrt{- g_{E}}\,\left(\tilde{R}_{E}+e^{-2\tau-\frac{2k}{d-2}\tau}\,\hat R_k-K_{\sigma\sigma}\,( \partial \,\sigma)^2-K_{\tau\tau}\,( \partial \,\tau)^2 \right)\,,
 \end{split}
\end{equation}
where 
\begin{equation}
 \begin{split}\label{metricdistanceKmt}
  &K_{\sigma\sigma}=-(d-1)(d-2)\,,\qquad \qquad K_{\tau\tau}=k^2\,\left(\frac{d-1}{d-2}\,-\frac{k-1}{k}\right)\,.
 \end{split}
\end{equation}

Within a family of AdS solutions in string theory, for example AdS$_4 \times S^7$ vacua, the external AdS radius and the internal volume are related. Our proposal for calculating the metric on such a family of solutions is that this on-shell relation in the family is promoted to hold also in the general `gauged' variations. We therefore impose a relation
\be
\tau = a \;\sigma \;,
\ee
with $a$ some constant. This then yields a total metric on the metric variations
\begin{equation}\label{generalmetricdistance}
K^{\text{metric}}=K_{\sigma\sigma}+K_{\tau\tau}=-(d-1)(d-2)+a^2k^2\,\left(\frac{d-1}{d-2}\,-\frac{k-1}{k}\right)\,.
\end{equation}

We do not expect this to be the complete metric on families of solutions, due to contributions from varying fluxes, dilaton, and other components of solutions. Nonetheless, it is already interesting to consider how this metric behaves for various families of solutions. Freund-Rubin vacua, and more generally any families for which the internal dimensions are the same scale as the external AdS radius, have $a=1$. Then one finds: 
\bea
\mrm{AdS}_4\times S^7 \;&:&\; K^{\text{metric}} = \frac{51}{2} \;, \nonumber \\
\mrm{AdS}_7\times S^4 \;&:&\; K^{\text{metric}} = -\frac{114}{5} \;, \nonumber \\
\mrm{AdS}_5\times S^5 \;&:&\; K^{\text{metric}} = \frac{4}{3} \;.
\eea
Generally, for $a=1$, the metric is positive in String (and M-theory) for $d < 6$, and negative for larger $d$. 

Vacua where there is separation of scales have $a<1$, so this leads to a more negative metric on the solutions. If we have strong separation of scale, $a \ll 1$, then the metric becomes negative always.

\section{The metric on Freund-Rubin AdS vacua}
\label{sec:metflux}

%


In the previous section we considered families of solutions where the AdS conformal factor is varied along with the internal volume. In string theory, or M-theory, such families also have variations in other fields. It was proposed in \cite{Lust:2019zwm} that the distance in the field space between solutions with different vacuum energy should be defined including the entire set of possible field variations and not only those ones from the internal geometry. This is motivated by the expectation that in the UV completion there should not be any privileged variation and all of them should ultimately contribute to the space of deformations of a given AdS vacua. 

In this section we consider the simplest families of solutions in the sense that  only one other field varies in the family. These are the Freund-Rubin vacua in M-theory, where the additional varying field is associated to the changing Freund-Rubin flux. 


\subsection{Freund-Rubin vacua in M-theory}

Freund-Rubin AdS vacua are direct product geometries AdS$_d\times Y_k$ with a flux filling entirely the AdS volume or the internal space. We denote these two possible flux configurations as electric and magnetic respectively.
Freund-Rubin vacua in M-theory constitute an ideal set-up to study flux variations since they feature a non-trivial, but simple, contribution of the flux to the action. Moreover, they are described by one-parameter families of solutions of the equations of motion of 11-dimensional supergravity. 

In this section we review a way to think about the Freund-Rubin solutions in terms of reducing to a lower-dimensional action. So we match the solution to properties of an action over the AdS factor, coming from integrating over the sphere. This is prerequisite for the analysis we perform in section \ref{sec:fluxvas}, where we will use the same method to extract lower-dimensional kinetic terms for the parameter of the solutions. 

Note that, because there is no separation of scales between the AdS radius and the sphere one, the  lower dimensional action is not an effective action. Instead, one can understand it in terms of a consistent truncation. When we consider flux variations, we will also only directly integrate over the internal dimensions, to extract lower-dimensional kinetic terms for the parameter of the solutions. This is valid irrespective of whether there is an associated effective or consistent full action. 


\subsubsection{$\mrm{AdS}_7\times S^4$ vacua}\label{AdS7section}

 Let us consider AdS$_7\times S^4$ geometries in 11-dimensional supergravity,
\begin{equation}
 \begin{split}\label{AdS7vac}
  &ds_{11}^2=e^{2\sigma}\left(d\hat s_7^2+\frac{1}{4}\, d\hat s^2_{4}   \right)\,,\\
  &  F_{4}=\frac38\,e^{3\sigma}\,\text{vol}_{4}\,,
 \end{split}
\end{equation}
where $d\hat s_7^2$ and $d\hat s^2_{4}$ are respectively the metrics of unitary AdS$_7$ and $S^4$, and $\text{vol}_{4}$ is the volume form of unitary $S^4$. This background constitutes a one-parameter family of solutions of the equations of motion of 11-dimensional supergravity defined by the (square) radius of AdS$_7$, which in this notations is given by $e^{2\sigma}$. We stress that at this level $\sigma$ is just a parameter. Since we are interested in off-shell deformations of metric and fluxes we can cast the above geometries in a fashion which is more suitable for this purpose. Let us introduce the three parameters $\sigma,\, \tau,\, \alpha$ such that
\begin{equation}
 \begin{split}\label{AdS7vacOnshell}
  &ds_{11}^2=e^{2\sigma}d\hat s_7^2+\frac{1}{4}e^{2\tau}\, d\hat s^2_{4} \,,\\
  &  F_7=6\, e^{\alpha}\,\text{vol}_7\qquad \text{with}\qquad    F_{4}=\star  F_{7}=\frac38\,e^{\alpha-7\sigma+4\tau}\,\text{vol}_{4}\,,
 \end{split}
\end{equation}
where $\text{vol}_{7}$ is the volume form of unitary AdS$_7$. The 11-dimensional equations of motion are thus satisfied by the following on-shell conditions
\begin{equation}\label{AdS7onshellconditions}
 \text{on-shell:}\qquad \qquad \sigma=\tau \qquad \qquad  \text{and}\qquad \qquad \alpha=6\,\sigma\,.
\end{equation}
It is easy to verify that imposing these conditions on \eqref{AdS7vacOnshell} one obtains \eqref{AdS7vac}.
We can now calculate an on-shell 7-dimensional action associated to \eqref{AdS7vacOnshell}. 
To this aim we can just evaluate the 11-dimensional supergravity action\footnote{We don't include the Chern-Simons term since it is trivial in our analysis.}
\begin{equation}\label{11daction}
 S=\frac{1}{2\kappa_{11}^2}\int d^{11}x\,\sqrt{-g}\,\left(R-\frac12\,|F_4|^2 \right)
\end{equation}
on the AdS$_7\times S^4$ backgrounds \eqref{AdS7vacOnshell}. We can thus derive a 7-dimensional action by writing the 7d metric in the Einstein frame as it follows $e^{2\sigma}d\hat s^2_7=e^{2\omega}ds_E^2=e^{2\sigma}\,e^{2\omega}d\hat s_E^2$ with $\omega=-\frac45\,\tau$. In this way we obtain the following action\footnote{We imposed that $\frac{4^{-2}\text{Vol}(S^4)}{\kappa_{11}^2}=1$ and $\sqrt{- g_E}=e^{7\sigma}\sqrt{- \hat g_E}$.}
\begin{equation}
\begin{split}\label{actioncc7d}
 S_7=\frac12\int d^7x\sqrt{- g_E}\,\left(e^{-2\sigma}\hat R_E+48\,e^{-2\,\tau+2\omega}-18 e^{2\alpha-14\sigma+2\omega}\right)\,,
 \end{split}
\end{equation}
where the second and third terms are respectively associated to the curvature of the 4-sphere and to the flux action.

If one imposes the on-shell conditions \eqref{AdS7onshellconditions}, the last two terms give the 7-dimensional cosmological constant 
\be
-2\Lambda=30\,e^{-2\sigma+2\omega} \;,
\ee
where we point out that the $e^{2\omega}$ factor appearing in $\Lambda$ is related to the AdS$_7$ radius in the Einstein frame, which is not unitary in our conventions.

\subsubsection{$\mrm{AdS}_4\times S^7$ vacua}\label{AdS4section}

Let us now focus on Freund-Rubin vacua in M-theory featured by a purely electric flux, namely AdS$_4\times S^7$. Similarly to AdS$_7\times S^4$ vacua, these constitute a one-parameter familiy of solutions of the equations of motion whose geometry is given by
\begin{equation}
 \begin{split}\label{AdS4vac}
  &ds_{11}^2=e^{2\sigma}\left(d\hat s_{4}^2+4\, d\hat s^2_{7}   \right)\,,\\
  &  F_{4}=-3\,e^{3\sigma}\,\text{vol}_{4}\,.
 \end{split}
\end{equation}
Now $d\hat s_4^2$ and $d\hat s^2_{7}$ are the metrics of unit AdS$_4$ and $S^7$, and the 4-form $\text{vol}_{4}$ represents the volume form of unitary AdS$_4$. We can express the vacuum geometry \eqref{AdS4vac} in terms of the three parameters $\sigma, \,\tau,\, \alpha$ as we did for magnetic vacua in the previous section,
\begin{equation}
 \begin{split}\label{AdS4vacOnshell}
  &ds_{11}^2=e^{2\sigma}\,d\hat s_{4}^2+4\,e^{2\tau}\, d\hat s^2_{7}\,,\\
  &  F_{4}=-3\,e^{\alpha}\,\text{vol}_{4}\,.
 \end{split}
\end{equation}
 One can verify that the equations of motion are solved by the following conditions
\begin{equation}\label{AdS4onshellconditions}
 \text{on-shell:}\qquad \qquad \sigma=\tau \qquad \qquad  \text{and}\qquad \qquad \alpha=3\,\sigma\,.
\end{equation}
 We would like to reduce the 11-dimensional action \eqref{11daction} over this vacuum geometry and extract the 4-dimensional cosmological constant. The procedure here is more subtle because 4-forms in four dimensions are top-forms. One then has to introduce certain boundary terms for them \cite{Duff:1989ah,Groh:2012tf}. 
 To be more concrete on this point, let us try to derive the on-shell action just by evaluating the 11-dimensional action over the backgrounds \eqref{AdS7vacOnshell},
 \begin{equation}
 \begin{split}\label{actioncc4dwrong}
 S_4=\frac12\int d^4x\sqrt{- g_E}\,\left(e^{-2\sigma}\hat R_E+\frac{21}{2}\,e^{-2\,\tau+2\omega}+\frac92\,e^{2\alpha-8\sigma+2\omega}\right)\,,
 \end{split}
\end{equation}
 where we casted the 4d metric\footnote{We imposed that $\frac{2^{7}\text{Vol}(S^7)}{\kappa_{11}^2}=1$ and $\sqrt{- g_E}=e^{4\sigma}\sqrt{- \hat g_E}$.} in the Einstein frame $e^{2\sigma}d\hat s_4^2=e^{2\omega}ds_E^2=e^{2\sigma}\,e^{2\omega}d\hat s_E^2$ with $\omega=-\frac72\,\tau$.
If one imposes the on-shell conditions \eqref{AdS4onshellconditions}, it is easy to see that one does not reproduce the value of the cosmological constant in four dimensions because of the ``wrong" sign in front of the flux term. 

This issue was addressed in \cite{Duff:1989ah} (see also \cite{Groh:2012tf,Andriot:2020lea} for more recent discussions) through the introduction of a suitable boundary term.\footnote{We thank Niccolò Cribiori for reminding us of this point.} Such a term does not modify the equations of motion, but it crucially reproduces a consistent on-shell action.
To derive this boundary contribution, let us consider the 4-dimensional action for a 4-form $ F_4$.
We first need to introduce a gauge potential $C_3$ such that $F_4=d C_3$, and second, derive the equations of motion for $C_3$. Taking the variations of the 4-dimensional action with respect to $C_3$ leads to the equations of motion, $d (\star_4  F_4)=0$, plus a boundary condition, $\delta  C_3|_{boundary}=0$. The latter is usually imposed as a physical prescription on the behaviour of fields at infinity.    
 However, as pointed out in \cite{Groh:2012tf,Andriot:2020lea}, this condition is not gauge invariant and then it does not reproduce a formally well-defined on-shell action. In other words, to be formally correct, one should complement the action with a boundary term which cancels exactly this contribution, once the action is evaluated on-shell. We can thus proceed by subtracting from the initial flux action the boundary term responsible for the gauge symmetry breaking. This leads to the following action
\begin{equation}
\begin{split}\label{boundaryaction4d}
 \tilde{S}_{4}&=-\frac12\,\int d^4x\sqrt{- g_4}\,| F_4|^2+\int d^4x\sqrt{- g_4}\,\text{d}(  C_3\wedge \star_4  F_4)\,.
 \end{split}
\end{equation}
If we evaluate the above action on-shell, namely imposing the equations of motion $d (\star_4  F_4)=0$, we obtain
\begin{equation}
 \tilde{S}_{4}|_{\text{on-shell}}=+\frac12\,\int d^4x\sqrt{- g_4}\,| F_4|^2.
\end{equation}
Crucially, the sign of the flux part in the action is flipped to positive.

We can now come back to our initial problem of calculating an on-shell action for AdS$_4\times S^7$ vacua. When we reduce the 11-dimensional action over the geometries \eqref{AdS7vacOnshell} we obtain a theory coupled to a 4-flux whose on-shell value is given in \eqref{AdS4onshellconditions}. If we redefine the 4-dimensional action including the boundary term as in \eqref{boundaryaction4d}, we obtain an expression which differs from \eqref{actioncc4dwrong} only for a crucial sign in front of the flux term,
 \begin{equation}
 \begin{split}\label{actioncc4d}
 \tilde S_4=\frac12\int d^4x\sqrt{- g_E}\,\left(e^{-2\sigma}\hat R_4+\frac{21}{2}\,e^{-2\,\tau+2\omega}-\frac92\,e^{2\alpha-8\sigma+2\omega}\right)\,.
 \end{split}
\end{equation}
 After this procedure we are finally able to extract the right value of the 4-dimensional vacuum energy. Imposing the on-shell conditions \eqref{AdS4onshellconditions}, we obtain 
 \be
 -2\Lambda=6\,e^{-2\sigma+2\omega} \;,
 \ee 
 where, as for AdS$_7\times S^4$ case, the $e^{2\omega}$ factor appearing in $\Lambda$ indicates that AdS$_4$ radius in the Einstein frame is not unitary in our conventions.
 
 \subsection{Flux variations}
 \label{sec:fluxvas}
 
 We are now ready to introduce flux variations for Freund-Rubin vacua. We will keep our analysis general considering Freund-Rubin fluxes in $D$-dimensional vacua of the type AdS$_d\times Y_k$. This will allow us to formulate a general prescription for flux variations holding both for electric fluxes (like in AdS$_4\times S^7$) and magnetic ones (like in AdS$_7\times S^4$).
 
 Our proposal consists of expressing the flux in terms of a gauge potential and then take variations of the latter. To this aim we can start by introducing an (undeformed) gauge potential $\hat C_p$ as
\begin{itemize}
 \item Electric flux: $\qquad \hat F_d=d\hat C_p\,$,
 \item Magnetic flux: $\qquad \hat F_k=\star \hat F_d\qquad$ with $\qquad \hat F_d=d\hat C_p\,$,
 \end{itemize}
 with $p=d-1$. Specifically, for AdS$_4\times S^7$ we write $\hat F_4=d\hat C_3$ , and for AdS$_7\times S^4$ one has $\hat F_7=d\hat C_6$ with $\hat F_4= \star \hat F_7$.
In order to be as explicit as possible, it is useful at this point to introduce a parametrization for AdS$_d$. We choose the Poincar\'e patch, which is the most suitable for our analysis,
\begin{equation}
\begin{split}\label{AdScoordPoincare}
 &d\hat s^2_{d}=z^{-2}\bigl(ds^2_{M_p}+dz^2\bigr),\\
 &\text{vol}_{d}=z^{-d}\,dx^0\wedge \cdots \wedge dx^{p-1}\wedge dz\,,
 \end{split}
\end{equation}
where $ds^2_{M_p}=-(dx^0)^2+\dots+(dx^{p-1})^2$ is the metric over the $p$-dimensional Minkowski spacetime $M_p$ and $\text{vol}_{d}$ the volume of unitary AdS$_d$ space. If we look at on-shell fluxes in \eqref{AdS7vacOnshell} and \eqref{AdS4vacOnshell}, we can introduce an explicit form for $\hat C_p$ as
\begin{equation}\label{underformedC}
 \hat C_p=-z^{-p}\,dx^0\wedge \cdots \wedge dx^{p-1}\qquad \text{with}\qquad \hat F_d=d\hat C_p=(-1)^p\,p\, \text{vol}_{d} \,.
\end{equation}
We can use these explicit expressions to rewrite the flux configurations for Freund-Rubin vacua \eqref{AdS7vacOnshell} and \eqref{AdS4vacOnshell} in a more compact form,
\begin{equation}
 \text{on-shell}:\qquad F_d=e^{\alpha}\,\hat F_d=(-1)^p\,e^{\alpha}\,p\, \text{vol}_{d} \,,
\end{equation}
where $\alpha$ is a constant parameter which in the cases $p=6$ and $p=3$ is fixed by the on-shell conditions \eqref{AdS7onshellconditions} and \eqref{AdS4onshellconditions} for AdS$_7\times S^4$ and AdS$_4\times S^7$.

To generate kinetic terms for the flux variations, we need to promote $\alpha$ to a function defined over AdS$_d$. To this aim, we introduce the off-shell gauge potential $C_p$ and field strength $F_d$ as 
\begin{equation}\label{deformedFd}
\begin{split}
&C_p=e^{\alpha}\hat C_p\qquad \text{with}\qquad F_d=d C_p=e^{\alpha}\,(\hat F_d+\,d\alpha\wedge \hat C_p)\,.
  \end{split}
 \end{equation}
 We stress that in the above prescription we are adopting the analogue logic that we applied to metric variations $\sigma,\,\tau$ in section \ref{sec:genactiondistance}, specifically, we do not need at this level to impose any particular restriction on $\alpha$. Nevertheless, because of the specific form of the undeformed gauge potential $\hat C_p$ in (\ref{underformedC}), we see that only derivatives along the $z$ direction contribute to the flux deformation. 
 Calculating explicitly the off-shell flux we obtain
 \begin{equation}\label{deformedFdexplicit}
  F_d=(-1)^p\,e^{\alpha}\,\left(p\,-z\,\partial_z\alpha\right) \text{vol}_{d}\,.
 \end{equation}
We now would like to insert this flux into the action, and extract lower dimensional kinetic terms for $\alpha$.

Note that we cannot treat the flux in (\ref{deformedFdexplicit}) as a true flux, for example, as appearing in the equations of motion. We are not satisfying the equations of motion, nor we are satisfying quantization conditions on the flux. For us, the flux (\ref{deformedFdexplicit}) is simply what must be inserted into the action to yield the kinetic terms for $\alpha$. Further, these kinetic terms will only appear with the derivative with respect to $z$, picking out this direction as special. This is related to the gauge choice for the 3-form (\ref{underformedC}). We could not find a fully covariant way to reach a well-defined distance contribution. There are yet further open questions: a top form flux is associated with a form field which is not dynamical, since its field strength is dual to a scalar. Therefore, we cannot expect to induce true dynamics for the parameter $\alpha$ from it, at least not on-shell. This is compatible with $\alpha$ not being a true dyanmical degree of freeedom in the theory, but only something from which one can extract a metric on flux variations. These are all conceptual problems which we are dismissing at this stage of the investigation. 

%


\subsubsection{Compensating metric variations}
\label{sigma1section}

Apart from the conceptual questions surrounding (\ref{deformedFdexplicit}), there is a technical issue which arises: in the action the square of (\ref{deformedFdexplicit}) appears. This means that apart from the term that is quadratic in derivatives acting on $\alpha$, there will be mixed terms, linear in derivatives, involving the background flux. It is not clear what to make of these mixed terms. We propose that they actually can be cancelled through an appropriate compensating metric variation. This is the topic of this section.

%

In choosing to promote the parameter $\sigma$ in (\ref{externalvariations}) from a constant to some function over AdS, what we called the gauging procedure, there is a certain ambiguity. Consider the following metric 
\begin{equation}\label{AdSdeformed}
\begin{split}
 &ds_{D}^2=e^{2\sigma} d\hat s^2_d+e^{2\tau}d\hat s^2_k\,,\\
 &d\hat s^2_d=\frac{1}{z^2}\left(ds^2_{M_p}+e^{2\sigma_1}dz^2\right) \;.
 \end{split}
\end{equation}
Taking both $\sigma$ and $\sigma_1$ as constants, the external part of the metric $ds^2_D$, so $e^{2\sigma} d\hat s^2_d$, describes AdS$_d$ with conformal factor $e^{2\sigma+2\sigma_1}$. This can be checked by performing the reparametrisation $\bar z=e^{\sigma_1}\,z$. Therefore, for constant $\sigma$ and $\sigma_1$, we can absorb $\sigma_1$ into $\sigma$ without loss of generality. However, once we gauge the parameters $\sigma$ and $\sigma_1$, so give them some spatial dependence, it is not longer true that $\sigma_1$ is identical to $\sigma$. Therefore, the general gauging procedure for the conformal factor is actually one where both $\sigma$ and $\sigma_1$ are gauged. 

If we gauge the parameter $\sigma_1$, then we should understand its possible contribution to the distance. 
We are therefore interested in computing the Einstein-Hilbert action for metric variations \eqref{AdSdeformed}. To this aim, we restrict to the case
\be
\sigma_1=\sigma_1(z)\;. 
\ee
The reason for this restriction is due to the particular form of the flux variations \eqref{deformedFdexplicit}, which contribute to the action only with derivatives of $\alpha$ along the $z$ direction.
In appendix \ref{AdSbreaking} we derive the Ricci scalar $\hat R_d$ associated to metric $d\hat s^2_d$ appearing in \eqref{AdSdeformed}. Here we write only the final expression
\begin{equation}\label{RicciscalarLinearTerm}
  \hat R_d=\hat{\tilde R}_d-2pe^{-2\sigma_1}z\,\partial_z\sigma_1 \,,
\end{equation}
with $\hat{\tilde R}_d=-d(d-1)e^{-2\sigma_1}$ and $p=d-1$.
From this expression we observe that the inclusion of the new modulus $\sigma_1$ has the unique effect to produce a new linear term in $\partial_z\sigma_1$ into the action for metric variations of AdS$_d$ space.

We can now follow the same procedure of section \ref{sec:genactiondistance}. The unique difference is represented by the Ricci scalar $\hat R_d$ which is now given by \eqref{RicciscalarLinearTerm} when we evaluate the action on deformed geometries \eqref{AdSdeformed}.
We perform the complete derivation of the Einstein-Hilbert action in appendix \ref{AdSbreaking} and we present here the final result,
\begin{equation}
\begin{split}\label{AdSmetricvariationaction}
 S_{\text{EH},d}=&\,\frac12\int d^dx\,\,\sqrt{- g_{E}}\bigl(e^{-2\sigma}\hat{\tilde R}_{E}-2pe^{-2\sigma-2\sigma_1+2\omega}z\,\partial_z\sigma_1+e^{-2\tau+2\omega}\hat R_k\\
 &-K_{\sigma\sigma}\,( \partial \,\sigma)^2-K_{\tau\tau}\,( \partial \,\tau)^2\bigr)\,,
 \end{split}
\end{equation}
where external volume variations have been absorbed within the (deformed) Einstein frame metric as $ds^2_{E}=e^{2\sigma}d\hat s^2_{E}$ and the components $K_{\sigma\sigma}$ and $K_{\tau\tau}$ are as in (\ref{metricdistanceKmt}).
  We point out that we kept completely explicit the $z$-dependence in the linear term in $\partial_z\sigma_1$ to make more manifest the cancellation occuring when flux variations are taken into account, as we are going to see in next section.
  
  \subsubsection{The metric over flux variations}\label{fluxvariations}
 
We are now ready to derive the contribution to action from flux variations. We will firstly develop the analysis for electric fluxes (such as AdS$_4\times S^7$) in generic $D$ dimensions and secondly we will discuss the case of magnetic fluxes (such as AdS$_7\times S^4$). As we are going to see the resulting off-shell actions for flux variations take the same form.

\subsubsection*{Electric fluxes}

Let us first summarize our prescription for field variations:
\begin{equation}\label{deformedFd1}
\begin{split}
&ds_{D}^2=e^{2\sigma} d\hat s^2_d+e^{2\tau}d\hat s^2_k\qquad \text{with}\qquad d\hat s^2_d=\frac{1}{z^2}\left(ds^2_{M_p}+e^{2\sigma_1}dz^2\right)\,,\\
&C_p=e^{\alpha}\hat C_p\qquad \text{with}\qquad F_d=d C_p=(-1)^p\,e^{\alpha}\,\left(p\,-z\,\partial_z\alpha\right) \text{vol}_{d}\,,
  \end{split}
 \end{equation}
where we assumed that the underformed flux $\hat F_d$ is purely electric, namely it fills entirely the AdS$_d$ space, $\hat F_d=d\,\hat C_p=(-1)^p\,p\, \text{vol}_{d}$ with $\hat C_p=-z^{-p}\,\text{vol}_{M_p}$\footnote{We use the notation $\text{vol}_{M_p}=dx^0\wedge \cdots \wedge dx^{p-1}$.}. We can start from the usual flux action $S_F=\frac{1}{2\kappa_D^2}\int d^Dx\sqrt{-g}\,(-\frac12|F_d|^2)$ and evaluate it on the off-shell variations \eqref{deformedFd1}. This leads to the following expression in $D$-dimensions
\begin{equation}
\begin{split}\label{actionvariedfluxes0}
 S_F=\frac{1}{2\kappa_D^2}\int d^D x \sqrt{-g}\,\bigl(-\frac{1}{2}\,e^{2\alpha}|\hat F_d|^2+\frac12\,e^{2\alpha-2d\sigma-2\sigma_1}z^2(\partial_z\alpha)^2-p\,e^{2\alpha-2d\sigma-2\sigma_1}\,z\partial_z\alpha   \bigr)\,,
\end{split}
\end{equation}
where we obtain $|\hat F_d|^2=-p^2e^{-2d\sigma-2\sigma_1}$ when we evaluate the action on the Freund-Rubin flux $\hat F_d=(-)^p\,p\, \text{vol}_{d}$. We immediately observe that in addition to the kinetic term for $\alpha$, the above expression includes a contribution which is linear in $\partial_z\alpha$. As we argued at the beginning of section \ref{sigma1section}, understanding this contribution in terms of metric over the space of flux variations is problematic. The presence of this term motivates the prescription we have chosen for metric variations in \eqref{AdSdeformed}, where we introduced the modulus $\sigma_1$. In fact from \eqref{AdSmetricvariationaction} it is manifest that a similar term in $\partial_z\sigma_1$ is produced from the Einstein-Hilbert action. In what follows we will discuss the conditions for which the two linear terms in $\partial_z\sigma_1$ and $\partial_z\alpha$ cancel each other\footnote{The two linear terms are of the same type. In fact if we rewrite the metric $d\hat s_d^2$ in \eqref{deformedFd1} using the notation of appendix \ref{AdSbreaking}, namely as $d\hat s_d^2=e^{2\bar\sigma}(ds^2_{M_p}+e^{2\sigma_1}dz^2)$, then the linear term in $\partial_z \alpha$ in \eqref{actionvariedfluxes0} takes the form $-\,p\,e^{2\alpha-2d\sigma-2\sigma_1}\,z\,\partial_{z}\alpha=\,p\,e^{2\alpha-2p\sigma}\,g_d^{zz}\partial_z\bar \sigma\,\partial_{z}\alpha$ with $e^{2\bar \sigma}=z^{-2}$.}. 

First we need to dimensionally reduce the above off-shell action and cast it into the Einstein frame. To reduce the action we just need to manifest internal volume variations as it follows
\begin{equation}
\begin{split}\label{actionvariedflux1}
 S_F=\frac{1}{2\kappa_D^2}\int d^Dx\sqrt{\hat g_k}\sqrt{-g_d}e^{k\tau}&\bigl(-\frac12\,e^{2\alpha}|\hat F_d|^2+\frac 12 \,e^{2\alpha-2p\sigma}g_d^{zz}\partial_{z}\alpha\,\partial_z\alpha\\
 &-\,p\,e^{2\alpha-2d\sigma-2\sigma_1}\,z\,\partial_{z}\alpha  \bigr)\;,
  \end{split}
\end{equation}
and cast the lower-dimensional action into the Einstein frame\footnote{Note that we could have also first gone to the Einstein frame, and then performed the flux variation.}. We can thus proceed by introducing the Einstein frame metric $ds^2_{d}=e^{2\omega}d s^2_{E}$ with $d s^2_{E}=e^{2\sigma}d\hat s^2_{E}$ and $\omega=-\frac{k}{d-2}\,\tau$. The procedure is trivial since there are no derivatives of the metric involved. In this way we obtain a $d$-dimensional action of the form\footnote{In the reduction we imposed $\frac{\text{Vol}(Y_k)}{\kappa_D^2}=1$.}
\begin{equation}
\begin{split}\label{actionvariedflux2}
 S_{F,d}=\frac{1}{2}\int d^dx\sqrt{-g_{E}}\,&\bigl(-\frac12\,e^{2\alpha-2p\omega}|\hat F_d|_E^2+\frac 12 \,e^{-2p\sigma+2\alpha}(\partial\alpha)^2_z\\
 &-\,p\,e^{2\omega-2\sigma-2\sigma_1+2\alpha-2p\sigma}\,z\,\partial_{z}\alpha  \bigr)\,,
  \end{split}
\end{equation}
where with $|\hat F_d|_E^2$ we indicate that the contractions are performed with respect to the Einstein frame metric. We have also introduced the notation
\be
(\partial\alpha)_z^2=g_{E}^{zz}\partial_{z}\alpha\,\partial_z\alpha \;.
\ee
So, from the above analysis we conclude that our prescription on fluxes leads to a new contribution to the distance over the space of field variations plus a linear term. 

We may conclude the analysis for electric fluxes by recalling an important fact regarding actions of electric vacua discussed in section \ref{AdS4section}. The dimensional reduction of electric fluxes defines top-forms in $d$-dimensions, and as we discussed in section \ref{AdS4section}, a boundary term is needed in order to reproduce a well-defined on-shell action. Such a boundary term crucially switchs the sign in front of the on-shell flux action (see equation \eqref{boundaryaction4d} and the subsequent discussion). The action for flux variations has to be consistent with this fact, otherwise we simply cannot obtain the right action when we go on-shell and set $\alpha$ to a constant. This means we should flip the sign of the off-shell action \eqref{actionvariedflux2} as $ S_{F,d}\rightarrow \tilde  S_{F,d}=- S_{F,d}$. We thus finally obtain
\begin{equation}
\begin{split}\label{actiondistance1}
 \tilde S_{F,d}=&\frac{1}{2}\int d^dx\,\sqrt{- g_{E}}\,\bigl(\,\frac12\,e^{2\alpha-2p\omega}|\hat F_d|_E^2-\frac12 \,e^{-2p\sigma+2\alpha}(\partial\alpha)^2_z+p\,e^{2\omega-2\sigma-2\sigma_1-2p\sigma+2\alpha}z\partial_{z}\alpha \bigr)\,.
 \end{split}
\end{equation}

\subsubsection*{Magnetic fluxes}

In the case of magnetic fluxes, we impose the flux variations on the electric dual flux $F_d$, and then we Hodge-dualize $F_k=\star F_d$ to obtain the deformed flux contributing to the off-shell action. 
The action can be thus worked out similarly to the previous case. In fact since the electric dual flux fills the AdS$_d$ part of the metric we can use again the prescriptions \eqref{deformedFd1}. However, this time we must evaluate the contribution to the action from $S_F=\frac{1}{2\kappa_D^2}\int d^Dx\sqrt{-g}\,(-\frac12|F_k|^2)$ with $F_k=\star F_d$. Specifically, we have the following expressions
\begin{equation}
\begin{split}
 &C_p=-e^{\alpha}z^{-p}\text{vol}_{M_p}\,,\qquad \qquad F_d=(-1)^p\,e^{\alpha}\,\left(p\,-z\,\partial_z\alpha\right) \text{vol}_{d}\,,\\
 &F_k=(-1)^{p}\,e^{\alpha-d\sigma-\sigma_1+k\tau}\,\left(p\,-z\,\partial_z\alpha\right) \text{vol}_{k}\;.
 \end{split}
\end{equation}
For instance, for AdS$_7\times S^4$ one has $C_6=-e^{\alpha}\,z^{-6}\,\text{vol}_{M_6}$ and $F_7=e^{\alpha}\,\left(6-z\partial_z\alpha  \right)\text{vol}_7$. The 4-flux is thus given by $F_4=\star\,F_7=\frac{1}{16}\,e^{\alpha-7\sigma-\sigma_1+4\tau}\,\left(6-z\partial_z\alpha  \right)\text{vol}_4$.

The flux action for magnetic flux variations $F_k$ is the same to that one for $F_d$, up to an important sign flip. This fact can be observed from the following equalities\footnote{We used that $\star F_k=\star(\star F_d)=-(-1)^{kd}F_d$.}
\be\label{dualaction}
- F_k\,\wedge  \star F_k=(-1)^{kd}\,F_k\, \wedge\, F_d=F_d\,\wedge \,F_k=F_d\wedge \star F_d\,.
 \ee
It therefore follows that once we have taken into account this overall sign difference, we can repeat exactly the same analysis for electric flux variations. This leads to the action \eqref{actionvariedflux2} with an overall minus sign, namely
\begin{equation}
\begin{split}\label{actiondistance1.1}
 S_{F,d}=&\frac12\int d^dx\sqrt{- g_{E}}\,\bigl(\frac12\,e^{2\alpha-2p\omega}|\hat F_d|_E^2-\frac12 \,e^{-2p\sigma+2\alpha}(\partial\alpha)_z^2+p\,e^{2\omega-2\sigma-2\sigma_1-2p\sigma+2\alpha}z\,\partial_{z}\alpha \bigr).
 \end{split}
\end{equation}
We thus observe that our final off-shell action for magnetic flux variations has the same form of the electric case \eqref{actiondistance1}.

\subsubsection{The full metric on AdS vacua}

At this point we are ready to add the action for flux variations \eqref{actiondistance1} (and the equivalent \eqref{actiondistance1.1}), to the Einstein-Hilbert action for metric variations written in \eqref{AdSmetricvariationaction}. We obtain the following expression
\begin{equation}
\begin{split}\label{actiondistance12}
 S_d=&\,\frac12\int d^dx\,\sqrt{- g_{E}}\,\bigl(e^{-2\sigma}\hat {\tilde R}_{E}+e^{-2\tau+2\omega}\,\hat R_k+\frac12\,e^{2\alpha-2p\omega}|\hat F_d|_E^2\\
  &-K_{\sigma\sigma}\,( \partial \,\sigma)^2-K_{\tau\tau}\,( \partial \,\tau)^2-\frac12 \,e^{-2p\sigma+2\alpha}(\partial\alpha)_z^2\\
 &-2\,p\,e^{2\omega-2\sigma-2\sigma_1}\,z\,\partial_z\,\sigma_1+\,p\,e^{2\omega-2\sigma-2\sigma_1-2p\sigma+2\alpha}\,z\,\partial_{z}\alpha \bigr)\,,
 \end{split}
\end{equation}
where we remind that $p=d-1$ and $\omega=-\frac{k}{d-2}\,\tau$, and $K_{\sigma\sigma}$ and $K_{\tau\tau}$ are as in \eqref{metricdistanceKmt}. We point out that above action describes both electric and magnetic flux variations. Specifically, in the case of magnetic fluxes, we rewrote the undeformed action in terms of the electric dual flux $\hat F_d$ using \eqref{dualaction}.

We stress that we have not imposed yet any relation among the variations $\sigma, \sigma_1, \tau, \alpha$. As discussed in section \ref{genvariationsmetric}, our prescription is to promote the relations of the constant parameters to hold also for the gauged (spatially dependent) parameters. Before considering the $\sigma_1$ variations, we had the constant parameter conditions \eqref{AdS4onshellconditions} and \eqref{AdS7onshellconditions} for $\sigma, \tau, \alpha$ for AdS$_4\times S^7$ and for AdS$_7\times S^4$. When $\sigma_1$ is a constant, only one combination of $\sigma$ and $\sigma_1$ is fixed by the equations of motion. In what follows we exploit this freedom to make a particular choice for $\sigma_1$ variation which cancels the linear derivative terms appearing in \eqref{actiondistance1}. 

Because in the action the derivatives of $\sigma_1$ and $\alpha$ are only in the $z$-direction, in order to be able to define appropriate relations between the parameters we restrict $\sigma$ and $\tau$ to be functions of $z$ only
\be
\sigma = \sigma(z) \;\;\;,\; \tau = \tau(z) \;. 
\ee
We then take the following relations between our variations (so for non-constant, or off-shell, parameters)
\begin{equation}\label{general-onshell}
 \sigma+\sigma_1=\tau\,,\qquad \qquad\alpha=p\,\sigma\,,
\end{equation}
with $p=d-1$.
There are two ways to motivate these relations. One is that they are the correct relations on-shell, so when the parameters are constant. As we showed after equation \eqref{AdSdeformed}, when we include $\sigma_1$-variations and we go on-shell we obtain again the AdS$_d$ metric but with square radius rescaled as $e^{2\sigma+2\sigma_1}$. It is thus easy to verify that the equations of motion for AdS$_4\times S^7$ and AdS$_7\times S^4$ are solved by \eqref{general-onshell} once we reparametrize the $z$-coordinate as $z\rightarrow \bar z=e^{\sigma_1} z$ in the on-shell metrics \eqref{AdS4vacOnshell} and \eqref{AdS7vacOnshell}.
The second way to motivate the above relation relies only on the off-shell consistency of the action \eqref{actiondistance1}. In fact the condition $\alpha=p\, \sigma$ guarantees that the flux action is well-defined at infinite distance in the space of variations and of the same order of the Einstein-Hilbert one. Moreover the relation $\sigma+\sigma_1=\tau$ is needed to keep intact the scaling $e^{-2\sigma-2\sigma_1}$ for the cosmological constant.

As we said, the conditions \eqref{general-onshell} does not fixed uniquely the field variations, they are two equations with four parameters. We thus has the freedom to impose an additional relation. Looking at the two linear terms in $\partial_z\sigma_1$ and $\partial_z\alpha$ we immediately notice that they cancel if we impose 
\begin{equation}\label{cancmixedterm}
 \sigma_1=\frac{\alpha}{2}\,,
\end{equation}
together with relations \eqref{general-onshell}. We thus have three relations among our field variations which can be solved together,
\begin{equation}
 \tau=\left(1+\frac p2\right)\sigma\,,\qquad \qquad \sigma_1=\frac{p}{2}\,\sigma\,,\qquad \qquad\alpha=p\,\sigma\,.
 \label{onshrefins1}
\end{equation}

We can finally evaluate our off-shell action \eqref{actiondistance12} on the above constraints among field variations. This leads to the action metric for Freund-Rubin vacua variations
\begin{equation}\label{finalactiondistance}
  S_d=\frac12\int d^dx\,\sqrt{- g_{E}}\,\bigl(e^{-2\sigma}\hat {\tilde R}_{E}+e^{-2\tau+2\omega}\,\hat R_k+\frac12\,e^{2\alpha-2p\omega}|\hat F_d|_E^2-K\,(\partial\sigma)_z^2 \bigr)\,.
\end{equation}
In \eqref{finalactiondistance}, we introduced the total metric $K$ on the space of field variations which has the explicit form
\begin{equation}\label{finalmetric}
 K=-(d-1)(d-2)+\frac{k^2}{4}\,\left(\frac{d-1}{d-2}\,-\frac{k-1}{k}\right)\left(d+1\right)^2+\frac12(d-1)^2\,.
\end{equation}
The flux variation has led to two differences relative to the metric for the metric variations (\ref{generalmetricdistance}). First, we have the additional last term in \eqref{finalmetric}, which comes directly from the flux terms. Second, there is an additional factor of $\frac{1}{4}\left(d+1\right)^2$ multiplying the internal metric variations, $K_{\tau\tau}$, which is due to the fact that the relation between $\sigma$ and $\tau$ is modified by the inclusion of $\sigma_1$, from (\ref{AdS7onshellconditions}), (\ref{AdS4onshellconditions}) to (\ref{onshrefins1}). 

Importantly, we note that the kinetic terms for $\sigma$ in (\ref{finalactiondistance}) are with derivatives only along $z$. So the action is not covariant. This can be tracked back to our choice for the three-form potential (\ref{underformedC}). We could not find a covariant way to implement the flux variations, and therefore we have extracted the metric through the radial direction of AdS. Note that, as discussed in the introduction, once the metric is extracted, it is meant to be used to measure distances along constant (so on-shell) variations of the parameters, in which there is no spatial dependence on AdS at all.

We can now restrict the general formula (\ref{finalmetric}) to the specific cases of AdS$_4\times S^7$ and AdS$_7\times S^4$. We find
\bea
\mrm{AdS}_4\times S^7 \;&:&\; K = \frac{1563}{8} \;, \nonumber \\
\mrm{AdS}_7\times S^4 \;&:&\; K = \frac{516}{5} \;.
\eea
The metrics on these solutions are now positive, and therefore yield well-defined distances.

Finally, we note that we can apply also the formula (\ref{finalmetric}) directly to AdS$_5\times S^5$ solutions of type IIB string theory. The derivation of (\ref{finalmetric}) is more subtle due to the self-duality condition of the flux, meaning that one cannot utilise an action directly. But the result should be the same, and in this case we obtain
\be
\mrm{AdS}_5\times S^5 \;:\; K = 116 \;.
\ee 
Other sets of solutions in string theory are more subtle, because they usually involve multiple parameters and forms varying.

\section{Summary}
\label{sec:dis}

In \cite{Lust:2019zwm}, the idea of a distance between AdS solutions in quantum gravity was raised in the context of the AdS Distance Conjecture. In this paper, we began some initial developments of this notion of distance. The work is very much exploratory. We introduced the action metric, as a metric on variations of the action within some set of solutions. More precisely, it is the factor multiplying the kinetic terms in the action that arise when the constant parameter of the set of solutions is promoted to have spatial dependence (a procedure we called ``gauging"). There are three primary types of contributions to such an action metric, coming from variations of the external conformal factor of AdS, variations of the internal conformal factor (the volume), and variations of the fluxes. We gave well-defined  prescriptions for evaluating these contributions. 

We used our prescriptions to calculate the action metrics on Freund-Rubin families of solutions, specifically AdS$_4 \times S^7$ and AdS$_7 \times S^4$ solutions of eleven-dimensional supergravity. Of particular interest are the signs of the action metrics. We found that the action metrics are positive for both these families. The total positive metrics had a negative contribution from the external conformal factor, but were made positive once the contributions from the internal dimensions as well as the fluxes were accounted for.  

The calculations of the contributions to the action metric from the external and internal conformal factors were fairly straightforward. Considering only these, we already encountered an interesting phenomenon: positive action metrics are related to the absence of scale-separation between the internal and external dimensions. This is because if the AdS scales much stronger than the internal volume, the positive internal dimensions contribution would lose out to the negative external conformal factor contribution. 

It is natural to consider the possibility that consistent sets of solutions in quantum gravity must have positive action metrics over them, perhaps relating to solving the conformal factor problem, and this in turn is related to the absence of scale separation in many solutions of string theory. Indeed, it could be that requiring positive action metrics is the more general and fundamental Swampland constraint: a Positive Metric Conjecture. Then the AdS Distance Conjecture may be an implication of it.

The flux contribution was much more difficult to try and understand. We did find a well-defined prescription, but this relied on explicitly treating the radial direction of AdS as special. Indeed, instead of extracting covariant kinetic terms for the flux parameter, we only extracted kinetic terms along the radial direction. The origin of this issue comes from the choice of gauge potential for the Freund-Rubin flux, which naturally picks out the radial direction. It may be that a different covariant prescription to account for the flux contribution can be developed, and would yield different results. Although we could not find such an alternative, our prescription was the only thing we could make work. 

As emphasised multiple times, this work is very much exploratory. There are many aspects of our calculation which, although well-defined and rigid, do not have a clear physical meaning. We consider it as first steps towards an explicit calculation of some sort of a metric over solutions of string theory. We hope that it may, at least, stimulate some more exploration of these ideas. 

\vskip 20pt
	\noindent {\bf Acknowledgements:} We would like to thank Ofer Aharony, Niccolò Cribiori and Giuseppe Dibitetto for useful comments and discussions. The work of EP and NP was supported by the Israel Science Foundation (grant No. 741/20). The work of YL, EP and NP was supported by the German Research Foundation through a German-Israeli Project Cooperation (DIP) grant ``Holography and the Swampland".

\newpage
\appendix

    \section{Derivation of the reduction formula}
    \label{app:reduction}

In this appendix we derive the reduction formula \eqref{Rdimreduction}.
Let us start with the metric
\begin{equation} \label{metric_start_Appdx}
ds_D^2=ds_d^2+e^{2\tau}d\hat s_k^2 \,,
\end{equation}
where $\tau$ depends only in the external coordinates.
We decompose the metric (\ref{metric_start_Appdx}) as 
\begin{equation} \label{metric_decomp1_Appdx}
ds_D^2=e^{2\tau}\left(e^{-2\tau}ds_d^2+d\hat s_k^2 \right) 
\equiv e^{2\tau}\left(d\overline{s}_d^2 +d\hat s_k^2 \right)
\equiv e^{2\tau} d\widetilde{s}_D^2 \,.
\end{equation}

With the conformal transformation (\ref{metric_decomp1_Appdx}), the Ricci scalar, $R$ of $ds_D^2$ is linked to the Ricci scalar, $\widetilde{R}$ of $d\widetilde{s}_D^2$ by
\begin{equation} \label{ricci_1_Appdx}
R = e^{-2\tau} \left[ \widetilde{R} - (D-1)(D-2) \widetilde{g}^{ab} (\partial_a \tau) (\partial_b\tau) -2(D-1) {\widetilde{\nabla}_D}^2\tau \right] \,.
\end{equation}
By definition, the labels $a$ and $b$ should run from all the spacetime indices, but the warp factor $\tau$ depends only in the external coordinates, so $a$ and $b$ actually label only the external coordinates. We will use $(m,n)$ to label the external coordinates.

The laplacian can be expressed through the Laplace-Beltrami formula
\begin{align}
{\widetilde{\nabla}_D}^2\tau &= \frac{1}{\sqrt{-\overline{g}_d}\sqrt{\hat g_k}} \partial_a \left( \sqrt{-\overline{g}_d}\sqrt{\hat g_k} \, \widetilde{g}^{ab}\, \partial_b\tau \right) \\
&= \frac{1}{\sqrt{-\overline{g}_d}} \partial_m \left( \sqrt{-\overline{g}_d} \, \widetilde{g}^{mn}\, \partial_n\tau \right)\\
&= \frac{1}{e^{-d\tau}\sqrt{-g_d}} \partial_m \left( e^{-(d-2)\tau} \sqrt{-g_d} \, g_d^{mn}\, \partial_n\tau \right) \\
&= e^{2\tau} \left[ \nabla_D^2\tau - (d-2)g_d^{mn}(\partial_m\tau)(\partial_n\tau) \right] \,.
\end{align}
Thus, the Ricci scalar in (\ref{ricci_1_Appdx}) can be rewritten in terms of $g_{mn}$ as
\begin{equation} \label{ricci_2_Appdx}
R= e^{-2\tau} \widetilde{R} + (d+k-1)(d-k-2) g_d^{mn}(\partial_m\tau)(\partial_n\tau) -2 (d+k-1) \nabla_D^2\tau \,.
\end{equation}
The decomposition
\begin{equation} \label{pure_product_Appdx}
d\widetilde{s}_D^2 = d\overline{s}_d^2 +d\hat s_k^2
\end{equation}
is that of a pure product of external and internal spaces, and not a warped product. Therefore, the Ricci scalar of the $D$-dimensional line element $d\widetilde{s}_D^2$ is the sum of that of the external and internal line elements, $d\overline{s}_d^2$ and $d\hat s_k^2$,
\begin{equation} \label{separation_Ricci_Appdx}
\widetilde{R} = \overline{R}_d + R_k \,.
\end{equation}
Here, $\widetilde{R}\equiv \widetilde{g}^{ab} (\widetilde{R})_{ab}$, whereas $\overline{R}_d \equiv \overline{g}_d^{a_1b_1}(\overline{R}_d)_{a_1b_1}$ and $R_k \equiv g_k^{a_2b_2} (R_k)_{a_2b_2}$ (where $a_1,b_1$ and $a_2,b_2$ run respectively from the external and internal indices).
The equality (\ref{separation_Ricci_Appdx}) is a well-known result in differential geometry, but one can also derive it in components.%
\footnote{Indeed, the fact we have a pure product (\ref{pure_product_Appdx}) implies that the Christoffel symbols, $\Gamma^a_{bc}$, with respect to the $D$-dimensional metric are non-zero only when they are of the form $\Gamma^{a_1}_{b_1c_1}$ or $\Gamma^{a_2}_{b_2c_2}$, that is to say without mixed indices between external and internal coordinates.
Then, 
\begin{align}
(\widetilde{R})_{a_1b_1} &=\partial_{c}\Gamma^{c}_{a_1b_1} -\partial_{a_1}\Gamma^{c}_{cb_1} + \Gamma^{c}_{a_1b_1}\Gamma^{d}_{cd} - \Gamma^{c}_{a_1d}\Gamma^{d}_{b_1c}\\
&=\partial_{c_1}\Gamma^{c_1}_{a_1b_1} -\partial_{a_1}\Gamma^{c_1}_{c_1b_1} + \Gamma^{c_1}_{a_1b_1}\Gamma^{d_1}_{c_1d_1} - \Gamma^{c_1}_{a_1d_1}\Gamma^{d_1}_{b_1c_1} =(\overline{R}_d)_{a_1b_1} \,,
\end{align}
and similarly for $(\widetilde{R}_D)_{a_2b_2}=(R_k)_{a_2b_2}$. We then deduce (\ref{separation_Ricci_Appdx}).%
}

We now use the conformal transformation of the $d$-dimensional spacetime $d\overline{s}_d^2=e^{-2\tau}ds_d^2$, to re-express the $d$-dimensional Ricci scalar as
\begin{equation} \label{Ricci_conformal2_Appdx}
\overline{R}_d = e^{2\tau} \left[ {R}_d - (d-1)(d-2) {g}_d^{mn} (\partial_m \tau) (\partial_n\tau) -2(d-1) \nabla_d^2\tau \right] \,.
\end{equation}
Plugging (\ref{separation_Ricci_Appdx}) and (\ref{Ricci_conformal2_Appdx}) into (\ref{ricci_2_Appdx}), we get
\begin{equation}
R= R_d + e^{-2\tau}R_k -k(k+1) {g}_d^{mn} (\partial_m \tau) (\partial_n\tau) - 2k \nabla_d^2\tau \,.
\end{equation}

\section{Combined external and internal variations}
\label{app:extandint}

In this appendix we obtain a general expression of the action metric which takes into account both external and internal volume variations. To this aim we consider a spacetime $M_d\times Y_k$ and metric variations of the type
\begin{equation}\label{sigmataumetric}
 ds^2_D=e^{2\sigma}\,d\hat s^2_d+e^{2\tau}\,d\hat s^2_k\,,
\end{equation}
 where we allow to vary both internal and external volumes through the moduli fields $\sigma$ and $\tau$ defined over the external spacetime $M_d$. We remind that, as in sections \ref{externalvariations} and \ref{internalvariations}, the hatted quantities don't depend on $\sigma$ and $\tau$.
 
Let's proceed by steps. First we use compactification formula \eqref{Rdimreduction} to reduce the Einstein-Hilbert action to $d$-dimensions. In this way we take extract the contribution to the metric over internal volume variations ($\tau$-dependence). Secondly we derive the contributions to the action metric of external metric variations ($\sigma$-dependence). Finally we cast the action in the $d$-dimensional Einstein frame. We can start with the action derived in \eqref{internaleh},
\begin{equation}\label{genvar0}
 S=\frac{1}{2\kappa_D^2}\int d^Dx\sqrt{\hat g_k}\,\sqrt{- g_d}\,e^{k\,\tau}\,\left( R_d+e^{-2\tau}\,\hat R_k+k(k-1)\, g_d^{mn} \partial_m \,\tau \partial_n\tau \right)\,,
\end{equation}
where we performed the partial integration of the laplacian $\nabla_d^2\tau$ expressed in the Laplace-Beltrami form \eqref{laplacian}. We point out that this time the laplacian is defined in terms of the deformed metric $ds_d^2=e^{2\sigma}d\hat s^2_d$.

We may now extract the contribution from external volume variations by expanding $R_d$ using formula \eqref{WeylTransform},
\begin{equation}
\begin{split}\label{genvar1}
 S=\frac{1}{2\kappa_D^2}\int d^Dx&\sqrt{\hat g_k}\,\sqrt{-\hat g_d}\,e^{k\,\tau}e^{(d-2)\sigma}\,\bigl(\hat R_d+e^{-2\tau+2\sigma}\,\hat R_k+k(k-1)\,\hat g_d^{mn} \partial_m \,\tau \partial_n\tau\\
 &-(d-1)(d-2)\,\hat g_d^{mn} \partial_m \,\sigma \partial_n\sigma -2(d-1)\hat\nabla_d^2\sigma\bigr)\,.
 \end{split}
\end{equation}
 We note that the integration by parts of the laplacian $\hat\nabla_d^2\,\sigma$ produces a mixed term due to the presence of the overall factor $e^{k\tau}$,
 \begin{equation}
\begin{split}\label{genvar2}
 S=\frac{1}{2\kappa_D^2}\int d^Dx&\sqrt{\hat g_k}\,\sqrt{-\hat g_d}\,e^{k\,\tau}e^{(d-2)\sigma}\,\bigl(\hat R_d+e^{-2\tau+2\sigma}\,\hat R_k+k(k-1)\,\hat g_d^{mn} \partial_m \,\tau \partial_n\tau\\
 &+(d-1)(d-2)\,\hat g_d^{mn} \partial_m \,\sigma \partial_n\sigma +2(d-1)k\hat g_d^{mn} \partial_m \,\tau \partial_n\sigma\bigr)\,,
 \end{split}
\end{equation}
 where in the last line we summed up the two kinetic terms in $\sigma$ (one of them produced by integration by parts).
 
Now we are ready to pass to the Einstein frame. First we introduce the $d$-dimensional (undeformed) metric $d\hat s_{E}^2$ defined as $d\hat s_d^2=e^{2\omega}d\hat s_{E}^2$ with $\omega=-\frac{k}{d-2}\tau$. Then we expand the Ricci scalar $\hat R_d$ using again formula \eqref{WeylTransform},
 \begin{equation}
\begin{split}\label{generalvariationEinsteinframe}
 S=\frac{1}{2\kappa_D^2}\int d^Dx&\sqrt{\hat g_k}\sqrt{-\hat g_{E}}e^{(d-2)\sigma}\bigl(\hat R_{E}+e^{-2\tau+2\sigma+2\omega}\hat R_k+k(k-1)\hat g_{E}^{mn} \partial_m \tau \partial_n\tau\\
 &+(d-1)(d-2)\,\hat g_{E}^{mn} \partial_m \,\sigma \partial_n\sigma +2(d-1)k\,\hat  g_{E}^{mn}\, \partial_m \,\tau \partial_n\sigma\\
 &-(d-1)(d-2)\,\hat g_{E}^{mn} \partial_m \,\omega \partial_n\,\omega-2(d-1)\hat\nabla_E^2\,\omega\bigr)\,,
 \end{split}
\end{equation}
where $\hat\nabla_E^2\,\omega$ is defined w.r.t. the Einstein frame metric $d\hat s_{E}^2$. At this point we can integrate by parts the laplacian $\hat\nabla_E^2\,\omega$ producing the mixed term $(d-1)(d-2)\hat g_{E}^{mn} \partial_m \sigma \partial_n\omega$. This term cancels exactly with the mixed term in $\partial \tau\,\partial\sigma$ in \eqref{generalvariationEinsteinframe} after having imposed that $\omega=-\frac{k}{d-2}\tau$. 

We can thus cast expression \eqref{generalvariationEinsteinframe} as a $d$-dimensional action and absorbe the $\sigma$-dependence in the Einstein frame metric introducing $ds^2_{E}=e^{2\sigma}d\hat s^2_{E}$. In this way we obtain the following action\footnote{In the reduction we imposed $\frac{\text{Vol}(Y_k)}{\kappa_D^2}=1$.}
\begin{equation}\label{genvariations2}
 \begin{split}
  S_d=\frac{1}{2}\int d^dx\,\,\sqrt{- g_{E}}\,\left(e^{-2\sigma}\hat R_{E}+e^{-2\tau+2\omega}\,\hat R_k-K_{\sigma\sigma}\,( \partial \,\sigma)^2-K_{\tau\tau}\,( \partial \,\tau)^2 \right)\,,
 \end{split}
\end{equation}
where 
\begin{equation}
 \begin{split}\label{metricdistanceK}
  &K_{\sigma\sigma}=-(d-1)(d-2)\,,\qquad \qquad K_{\tau\tau}=k^2\,\left(\frac{d-1}{d-2}\,-\frac{k-1}{k}\right)\,.
 \end{split}
\end{equation}
 We point out that the action \eqref{genvariations2} could have been obtained by following a different order in the steps outlined above. In other words, external volume variations and internal variations plus the Einstein frame rescaling do commute. Specifically, from \eqref{genvar0} one can pass directly to the Einstein frame and then consider external metric variations. Such procedure is equivalent to derive the action metric using the following order for Weyl rescalings $ds^2_d=e^{2\omega}ds^2_{E}=e^{2\omega}e^{2\sigma}d\hat s^2_{E}$.

\section{The action of compensating metric variations}\label{AdSbreaking}

In this appendix we derive the Einstein-Hilbert action starting from the following deformations of the AdS$_d$ metric,
\begin{equation}
 d\hat s_d^2=\frac{1}{z^2}\bigl(ds^2_{M_p}+e^{2\sigma_1}\,dz^2\bigr)\,,
\end{equation}
where $ds^2_{M_p}=-(dx^0)^2+\dots+(dx^{p-1})^2$ and $\sigma_1=\sigma_1(z)$.
It is useful to cast the above metric in the slightly more general form $ d\hat s_d^2=e^{2\bar \sigma\,}d\bar s^2_d$ where $d\bar s^2_d=e^{2\sigma_1}dz^2+ds^2_{M_p}$ and $e^{2\bar\sigma}=z^{-2}$.
To calculate the Ricci scalar $\hat R_d$ we can use formula \eqref{WeylTransform} obtaining
\begin{equation}
\begin{split}\label{WeylMinkowski}
 \hat R_d=&e^{-2\,\bar \sigma}\,\left[\hat{\bar R}_d-(d-1)(d-2)\,\hat{\bar g}_d^{zz}\,(\partial_{ z}\bar\sigma)^2-2(d-1)\, \hat{\bar{\nabla}}_{d}^2\,\bar\sigma   \right]=\\
 &=-e^{-2\,\bar \sigma-2\sigma_1}\,\left[(d-1)(d-2)(\partial_z\bar\sigma)^2+2(d-1)\partial_{z}^2\bar\sigma\right]\\
 &+2(d-1)e^{-2\bar\sigma-2\sigma_1}\partial_z\bar\sigma\,\partial_z\sigma_1=\hat{\tilde R}_d+2(d-1)\hat g^{zz}_d\,\partial_z\bar\sigma\,\partial_z\sigma_1\,,
 \end{split}
\end{equation}
where $\hat{g}_d^{zz}=e^{-2\bar \sigma}\,\hat{\bar g}_d^{zz}=e^{-2\bar \sigma}e^{-2\sigma_1}$ and we used the Laplace-Beltrami laplacian given in \eqref{laplacian}. We also used that the Ricci scalar $\hat{\bar R}_d$ vanishes since $d\bar s^2_d$ describes a flat space. 
If we now specify the above expression to AdS$_d$, namely $e^{2\bar \sigma}=z^{-2}$, we obtain
\begin{equation}\label{ricciDW}
  \hat R_d=\hat{\tilde R}_d-2(d-1)e^{-2\sigma_1}z\,\partial_z\sigma_1 \,,
\end{equation}
where $\hat{\tilde R}_d=-d(d-1)e^{-2\sigma_1}$.
We can now derive the $d$-dimensional Einstein-Hilbert action for the following familiy of geometries
\begin{equation}
\begin{split}
  &ds^2_D=e^{2\sigma}d\hat s^2_d+e^{2\tau}d\hat s^2_k\,,\\
  &d\hat s_d^2=e^{2\bar\sigma}d\bar s^2_d\qquad \text{with}\qquad d\bar s_d^2=e^{2\sigma_1}dz^2+ds^2_{M_p}\,,
  \end{split}
\end{equation}
where $\bar\sigma$ and $\sigma_1$ are functions of $z$.
The action for the $\sigma$- and $\tau$-variations is given in \eqref{genvar2} and we may recall it here
\begin{equation}
\begin{split}\label{app:genvar2}
 S_{\text{EH}}=&\frac{1}{2\kappa_D^2}\int d^Dx\sqrt{\hat g_k}\,\sqrt{-\hat g_d}\,e^{k\,\tau}e^{(d-2)\sigma}\,\bigl(\hat{\tilde R}_d+2(d-1)\hat g^{zz}_d\,\partial_z\bar\sigma\,\partial_z\sigma_1+e^{-2\tau+2\sigma}\,\hat R_k\\
 &+k(k-1)\,\hat g_d^{mn} \partial_m \,\tau \partial_n\tau+(d-1)(d-2)\,\hat g_d^{mn} \partial_m \,\sigma \partial_n\sigma +2(d-1)k\,\hat g_d^{mn} \partial_m \,\tau \partial_n\sigma\bigr),
 \end{split}
\end{equation}
where in the above expression we use $\hat R_d=\hat{\tilde R}_d+2(d-1)\hat g^{zz}_d\,\partial_z\bar\sigma\,\partial_z\sigma_1$ derived in \eqref{WeylMinkowski} to extract the derivative term in $\partial_z\sigma_1$. We may pass to the Einstein frame through the Weyl rescaling $d\hat s_d^2=e^{2\omega}\,d\hat s^2_{E}$ with $\omega=-\frac{k}{d-2}\,\tau$,
\begin{equation}
\begin{split}\label{app:generalvariationEinsteinframe}
 S_{\text{EH}}=&\frac{1}{2\kappa_D^2}\int d^Dx\sqrt{\hat g_k}\sqrt{-\hat g_{E}}e^{(d-2)\sigma}\bigl(\hat{\tilde R}_{E}+2(d-1)\hat g^{zz}_{E}\,\partial_z\bar\sigma\,\partial_z\sigma_1+e^{-2\tau+2\sigma+2\omega}\hat R_k\\
 &+k(k-1)\hat g_{E}^{mn} \partial_m \tau \partial_n\tau+(d-1)(d-2)\,\hat g_{E}^{mn} \partial_m \,\sigma \partial_n\sigma +2(d-1)k\,\hat  g_{E}^{mn}\, \partial_m \,\tau \partial_n\sigma\\
 &-(d-1)(d-2)\,\hat g_{E}^{mn} \partial_m \,\omega \partial_n\,\omega-2(d-1)\hat\nabla_E^2\,\omega\bigr)\,,
 \end{split}
\end{equation}
where $\hat\nabla_E^2\,\omega$ is defined over the metric $d\hat s_{E}^2$. 
 As explained after equation \eqref{generalvariationEinsteinframe}, the integration by part of $\hat\nabla_E^2\omega$ cancels exactly the mixed term $2(d-1)k\,\hat  g_{E}^{mn}\, \partial_m \tau\, \partial_n\sigma$. Finally, we can evaluate the above action on the (deformed) AdS$_d$ geometry by specifying $e^{2\bar \sigma}=z^{-2}$. We thus obtain the following $d$-dimensional action\footnote{In the reduction we imposed $\frac{\text{Vol}(Y_k)}{\kappa_D^2}=1$.}
\begin{equation}
\begin{split}\label{app:generalvariationEinsteinframe}
 S_{\text{EH},d}=&\frac12\int d^dx\,\,\sqrt{- g_{E}}\bigl(e^{-2\sigma}\hat{\tilde R}_{E}-2pe^{-2\sigma-2\sigma_1+2\omega}z\,\partial_z\sigma_1+e^{-2\tau+2\omega}\hat R_k\\
 &-K_{\sigma\sigma}\,( \partial \,\sigma)^2-K_{\tau\tau}\,( \partial \,\tau)^2\bigr)\,,
 \end{split}
\end{equation}
where $p=d-1$ and we absorbed back external volume variations into the metric defining $ds^2_{E}=e^{2\sigma}d\hat s^2_{E}$. The metric components $K_{\sigma\sigma}$ and $K_{\tau\tau}$ are given in \eqref{metricdistanceK}. 

	\bibliographystyle{jhep}
	\bibliography{metricdistance}

\providecommand{\href}[2]{#2}\begingroup\raggedright\begin{thebibliography}{10}

\bibitem{Ooguri:2006in}
H.~Ooguri and C.~Vafa, \emph{{On the Geometry of the String Landscape and the
  Swampland}},
  \href{https://doi.org/10.1016/j.nuclphysb.2006.10.033}{\emph{Nucl. Phys. B}
  {\bfseries 766} (2007) 21}
  [\href{https://arxiv.org/abs/hep-th/0605264}{{\ttfamily hep-th/0605264}}].

\bibitem{Palti:2019pca}
E.~Palti, \emph{{The Swampland: Introduction and Review}},
  \href{https://doi.org/10.1002/prop.201900037}{\emph{Fortsch. Phys.}
  {\bfseries 67} (2019) 1900037}
  [\href{https://arxiv.org/abs/1903.06239}{{\ttfamily 1903.06239}}].

\bibitem{vanBeest:2021lhn}
M.~van Beest, J.~Calder\'on-Infante, D.~Mirfendereski and I.~Valenzuela,
  \emph{{Lectures on the Swampland Program in String Compactifications}},
  \href{https://doi.org/10.1016/j.physrep.2022.09.002}{\emph{Phys. Rept.}
  {\bfseries 989} (2022) 1} [\href{https://arxiv.org/abs/2102.01111}{{\ttfamily
  2102.01111}}].

\bibitem{Baume:2016psm}
F.~Baume and E.~Palti, \emph{{Backreacted Axion Field Ranges in String
  Theory}}, \href{https://doi.org/10.1007/JHEP08(2016)043}{\emph{JHEP}
  {\bfseries 08} (2016) 043}
  [\href{https://arxiv.org/abs/1602.06517}{{\ttfamily 1602.06517}}].

\bibitem{Klaewer:2016kiy}
D.~Klaewer and E.~Palti, \emph{{Super-Planckian Spatial Field Variations and
  Quantum Gravity}}, \href{https://doi.org/10.1007/JHEP01(2017)088}{\emph{JHEP}
  {\bfseries 01} (2017) 088}
  [\href{https://arxiv.org/abs/1610.00010}{{\ttfamily 1610.00010}}].

\bibitem{Blumenhagen:2018nts}
R.~Blumenhagen, D.~Kl\"awer, L.~Schlechter and F.~Wolf, \emph{{The Refined
  Swampland Distance Conjecture in Calabi-Yau Moduli Spaces}},
  \href{https://doi.org/10.1007/JHEP06(2018)052}{\emph{JHEP} {\bfseries 06}
  (2018) 052} [\href{https://arxiv.org/abs/1803.04989}{{\ttfamily
  1803.04989}}].

\bibitem{Grimm:2018ohb}
T.W.~Grimm, E.~Palti and I.~Valenzuela, \emph{{Infinite Distances in Field
  Space and Massless Towers of States}},
  \href{https://doi.org/10.1007/JHEP08(2018)143}{\emph{JHEP} {\bfseries 08}
  (2018) 143} [\href{https://arxiv.org/abs/1802.08264}{{\ttfamily
  1802.08264}}].

\bibitem{Grimm:2018cpv}
T.W.~Grimm, C.~Li and E.~Palti, \emph{{Infinite Distance Networks in Field
  Space and Charge Orbits}},
  \href{https://doi.org/10.1007/JHEP03(2019)016}{\emph{JHEP} {\bfseries 03}
  (2019) 016} [\href{https://arxiv.org/abs/1811.02571}{{\ttfamily
  1811.02571}}].

\bibitem{Corvilain:2018lgw}
P.~Corvilain, T.W.~Grimm and I.~Valenzuela, \emph{{The Swampland Distance
  Conjecture for K\"ahler moduli}},
  \href{https://doi.org/10.1007/JHEP08(2019)075}{\emph{JHEP} {\bfseries 08}
  (2019) 075} [\href{https://arxiv.org/abs/1812.07548}{{\ttfamily
  1812.07548}}].

\bibitem{Lee:2018spm}
S.-J.~Lee, W.~Lerche and T.~Weigand, \emph{{A Stringy Test of the Scalar Weak
  Gravity Conjecture}},
  \href{https://doi.org/10.1016/j.nuclphysb.2018.11.001}{\emph{Nucl. Phys. B}
  {\bfseries 938} (2019) 321}
  [\href{https://arxiv.org/abs/1810.05169}{{\ttfamily 1810.05169}}].

\bibitem{Lee:2019wij}
S.-J.~Lee, W.~Lerche and T.~Weigand, \emph{{Emergent strings from infinite
  distance limits}}, \href{https://doi.org/10.1007/JHEP02(2022)190}{\emph{JHEP}
  {\bfseries 02} (2022) 190}
  [\href{https://arxiv.org/abs/1910.01135}{{\ttfamily 1910.01135}}].

\bibitem{Lust:2019zwm}
D.~L\"ust, E.~Palti and C.~Vafa, \emph{{AdS and the Swampland}},
  \href{https://doi.org/10.1016/j.physletb.2019.134867}{\emph{Phys. Lett. B}
  {\bfseries 797} (2019) 134867}
  [\href{https://arxiv.org/abs/1906.05225}{{\ttfamily 1906.05225}}].

\bibitem{DeWitt}
B.S.~DeWitt, \emph{Quantum theory of gravity. i. the canonical theory},
  \href{https://doi.org/10.1103/PhysRev.160.1113}{\emph{Phys. Rev.} {\bfseries
  160} (1967) 1113}.

\bibitem{CANDELAS1991455}
P.~Candelas and X.C.~{de la Ossa}, \emph{Moduli space of calabi-yau manifolds},
  \href{https://doi.org/https://doi.org/10.1016/0550-3213(91)90122-E}{\emph{Nuclear
  Physics B} {\bfseries 355} (1991) 455}.

\bibitem{Gibbons:1976ue}
G.W.~Gibbons and S.W.~Hawking, \emph{{Action Integrals and Partition Functions
  in Quantum Gravity}},
  \href{https://doi.org/10.1103/PhysRevD.15.2752}{\emph{Phys. Rev. D}
  {\bfseries 15} (1977) 2752}.

\bibitem{Marolf:2022ybi}
D.~Marolf, \emph{{Gravitational thermodynamics without the conformal factor
  problem: partition functions and Euclidean saddles from Lorentzian path
  integrals}}, \href{https://doi.org/10.1007/JHEP07(2022)108}{\emph{JHEP}
  {\bfseries 07} (2022) 108}
  [\href{https://arxiv.org/abs/2203.07421}{{\ttfamily 2203.07421}}].

\bibitem{Rudelius:2021oaz}
T.~Rudelius, \emph{{Dimensional reduction and (Anti) de Sitter bounds}},
  \href{https://doi.org/10.1007/JHEP08(2021)041}{\emph{JHEP} {\bfseries 08}
  (2021) 041} [\href{https://arxiv.org/abs/2101.11617}{{\ttfamily
  2101.11617}}].

\bibitem{Basile:2021mkd}
I.~Basile, \emph{{Supersymmetry breaking, brane dynamics and Swampland
  conjectures}}, \href{https://doi.org/10.1007/JHEP10(2021)080}{\emph{JHEP}
  {\bfseries 10} (2021) 080}
  [\href{https://arxiv.org/abs/2106.04574}{{\ttfamily 2106.04574}}].

\bibitem{Angius:2022aeq}
R.~Angius, J.~Calder\'on-Infante, M.~Delgado, J.~Huertas and A.M.~Uranga,
  \emph{{At the end of the world: Local Dynamical Cobordism}},
  \href{https://doi.org/10.1007/JHEP06(2022)142}{\emph{JHEP} {\bfseries 06}
  (2022) 142} [\href{https://arxiv.org/abs/2203.11240}{{\ttfamily
  2203.11240}}].

\bibitem{Montero:2022ghl}
M.~Montero, M.~Rocek and C.~Vafa, \emph{{Pure supersymmetric AdS and the
  Swampland}}, \href{https://doi.org/10.1007/JHEP01(2023)094}{\emph{JHEP}
  {\bfseries 01} (2023) 094}
  [\href{https://arxiv.org/abs/2212.01697}{{\ttfamily 2212.01697}}].

\bibitem{Farakos:2023nms}
F.~Farakos, M.~Morittu and G.~Tringas, \emph{{On/off scale separation}},
  \href{https://arxiv.org/abs/2304.14372}{{\ttfamily 2304.14372}}.

\bibitem{Buratti:2020kda}
G.~Buratti, J.~Calderon, A.~Mininno and A.M.~Uranga, \emph{{Discrete
  Symmetries, Weak Coupling Conjecture and Scale Separation in AdS Vacua}},
  \href{https://doi.org/10.1007/JHEP06(2020)083}{\emph{JHEP} {\bfseries 06}
  (2020) 083} [\href{https://arxiv.org/abs/2003.09740}{{\ttfamily
  2003.09740}}].

\bibitem{Luben:2020wix}
M.~L\"uben, D.~L\"ust and A.R.~Metidieri, \emph{{The Black Hole Entropy
  Distance Conjecture and Black Hole Evaporation}},
  \href{https://doi.org/10.1002/prop.202000130}{\emph{Fortsch. Phys.}
  {\bfseries 69} (2021) 2000130}
  [\href{https://arxiv.org/abs/2011.12331}{{\ttfamily 2011.12331}}].

\bibitem{Li:2021utg}
Y.~Li, \emph{{An Alliance in the Tripartite Conflict over Moduli Space}},
  \href{https://arxiv.org/abs/2112.03281}{{\ttfamily 2112.03281}}.

\bibitem{Collins:2022nux}
T.C.~Collins, D.~Jafferis, C.~Vafa, K.~Xu and S.-T.~Yau, \emph{{On Upper Bounds
  in Dimension Gaps of CFT's}},
  \href{https://arxiv.org/abs/2201.03660}{{\ttfamily 2201.03660}}.

\bibitem{Cribiori:2022trc}
N.~Cribiori and G.~Dall'Agata, \emph{{Weak gravity versus scale separation}},
  \href{https://doi.org/10.1007/JHEP06(2022)006}{\emph{JHEP} {\bfseries 06}
  (2022) 006} [\href{https://arxiv.org/abs/2203.05559}{{\ttfamily
  2203.05559}}].

\bibitem{Shiu:2022oti}
G.~Shiu, F.~Tonioni, V.~Van~Hemelryck and T.~Van~Riet, \emph{{AdS scale
  separation and the distance conjecture}},
  \href{https://arxiv.org/abs/2212.06169}{{\ttfamily 2212.06169}}.

\bibitem{Cribiori:2023swd}
N.~Cribiori, A.~Gnecchi, D.~L\"ust and M.~Scalisi, \emph{{On the correspondence
  between black holes, domain walls and fluxes}},
  \href{https://doi.org/10.1007/JHEP05(2023)033}{\emph{JHEP} {\bfseries 05}
  (2023) 033} [\href{https://arxiv.org/abs/2302.03054}{{\ttfamily
  2302.03054}}].

\bibitem{Duff:1989ah}
M.J.~Duff, \emph{{The Cosmological Constant Is Possibly Zero, but the Proof Is
  Probably Wrong}},
  \href{https://doi.org/10.1016/0370-2693(89)90284-0}{\emph{Phys. Lett. B}
  {\bfseries 226} (1989) 36}.

\bibitem{Groh:2012tf}
K.~Groh, J.~Louis and J.~Sommerfeld, \emph{{Duality and Couplings of
  3-Form-Multiplets in N=1 Supersymmetry}},
  \href{https://doi.org/10.1007/JHEP05(2013)001}{\emph{JHEP} {\bfseries 05}
  (2013) 001} [\href{https://arxiv.org/abs/1212.4639}{{\ttfamily 1212.4639}}].

\bibitem{Andriot:2020lea}
D.~Andriot, N.~Cribiori and D.~Erkinger, \emph{{The web of swampland
  conjectures and the TCC bound}},
  \href{https://doi.org/10.1007/JHEP07(2020)162}{\emph{JHEP} {\bfseries 07}
  (2020) 162} [\href{https://arxiv.org/abs/2004.00030}{{\ttfamily
  2004.00030}}].

\end{thebibliography}\endgroup
\end{document}